\documentclass[useAMS,usenatbib]{mn2e}
\usepackage{graphicx,fleqn,fix2col,rotating}
\DeclareGraphicsExtensions{.eps,.ps,.eps.gz,.ps.gz,.eps.Z}
\title[SDSS J0926+3624: The shortest period eclipsing binary star]{SDSS J0926+3624: The shortest period eclipsing binary star}

\author[C.M.~Copperwheat et al.]{C.M.~Copperwheat$^{1}$, T.R.~Marsh$^{1}$,  S.P.~Littlefair$^{2}$, V.S.~Dhillon$^{2}$, G.~Ramsay$^{3}$, \newauthor A.J.~Drake$^{4}$, B.T.~G{\"{a}}nsicke$^{1}$, P.J.~Groot$^{5}$, P.~Hakala$^{6}$, D.~Koester$^{7}$, G.~Nelemans$^{5}$, \newauthor G.~Roelofs$^{8}$, J.~Southworth$^{9}$, D.~Steeghs$^{1}$ and S.~Tulloch$^{2}$\\\\
$^{1}$ Department of Physics, University of Warwick, Coventry, CV4 7AL, UK\\
$^{2}$ Department of Physics and Astronomy, University of Sheffield, S3 7RH, UK\\
$^{3}$ Armagh Observatory, College Hill, Armagh, BT61 9DG, UK\\
$^{4}$ California Institute of Technology, 1200 E. California Blvd., CA 91225, USA\\
$^{5}$ Department of Astrophysics, IMAPP, Radboud University Nijmegen, PO Box 9010, NL-6500 GL Nijmegen, the Netherlands\\
$^{6}$ Finnish Centre for Astronomy with ESO, Tuorla Observatory, V\"ais\"al\"antie 20, FIN-21500 Piikki\"o, University of Turku, Finland\\
$^{7}$ Institut f\"ur Theoretische Physik und Astrophysik, Universit\"at Kiel, 24098 Kiel, Germany\\
$^{8}$ Harvard-Smithsonian Center for Astrophysics, 60 Garden Street, Cambridge, MA 02138, USA\\
$^{9}$ Astrophysics Group, Keele University, Newcastle-under-Lyme, ST5 5BG, UK\\
}

\date{Received: }

\begin{document}

\newcommand{\dg} {^{\circ}}
\outer\def\gtae {$\buildrel {\lower3pt\hbox{$>$}} \over
{\lower2pt\hbox{$\sim$}} $}
\outer\def\ltae {$\buildrel {\lower3pt\hbox{$<$}} \over
{\lower2pt\hbox{$\sim$}} $}
\newcommand{\ergscm} {erg s$^{-1}$ cm$^{-2}$}
\newcommand{\ergss} {erg s$^{-1}$}
\newcommand{\ergsd} {erg s$^{-1}$ $d^{2}_{100}$}
\newcommand{\pcmsq} {cm$^{-2}$}
\newcommand{\ros} {{\it ROSAT}}
\newcommand{\xmm} {\mbox{{\it XMM-Newton}}}
\newcommand{\exo} {{\it EXOSAT}}
\newcommand{\sax} {{\it BeppoSAX}}
\newcommand{\chandra} {{\it Chandra}}
\newcommand{\hst} {{\it HST}}
\newcommand{\subaru} {{\it Subaru}}
\def\rchi{{${\chi}_{\nu}^{2}$}}
\newcommand{\Msun} {$M_{\odot}$}
\newcommand{\Mwd} {$M_{wd}$}
\newcommand{\Mbh} {$M_{\bullet}$}
\newcommand{\Lsun} {$L_{\odot}$}
\newcommand{\Rsun} {$R_{\odot}$}
\newcommand{\Zsun} {$Z_{\odot}$}
\def\Mdot{\hbox{$\dot M$}}
\def\mdot{\hbox{$\dot m$}}
\def\mincir{\raise -2.truept\hbox{\rlap{\hbox{$\sim$}}\raise5.truept
\hbox{$<$}\ }}
\def\magcir{\raise -4.truept\hbox{\rlap{\hbox{$\sim$}}\raise5.truept
\hbox{$>$}\ }}
\newcommand{\mnras} {MNRAS}
\newcommand{\aap} {A\&A}
\newcommand{\apj} {ApJ}
\newcommand{\apjl} {ApJL}
\newcommand{\apjs} {ApJS}
\newcommand{\aj} {AJ}
\newcommand{\pasp} {PASP}
\newcommand{\apss} {Ap\&SS}
\newcommand{\araa} {ARAA}
\newcommand{\nat} {Nature}
\newcommand{\pasj} {PASJ}
\maketitle

\begin{abstract} 
With orbital periods of the order of tens of minutes or less, the AM Canum Venaticorum stars are ultracompact, hydrogen deficient binaries with the shortest periods of any binary subclass, and are expected to be among the strongest gravitational wave sources in the sky. To date, the only known eclipsing source of this type is the $P = 28$ min binary SDSS~J0926+3624. We present multiband, high time resolution light curves of this system, collected with WHT/ULTRACAM in 2006 and 2009. We supplement these data with additional observations made with LT/RISE, \xmm \ and the Catalina Real-Time Transient Survey. From light curve models we determine the mass ratio to be $q = M_2 / M_1 = 0.041 \pm 0.002$ and the inclination to be $82.6 \pm 0.3$ deg. We calculate the mass of the primary white dwarf to be $0.85 \pm 0.04$\Msun \ and the donor to be $0.035 \pm 0.003$\Msun, implying a partially degenerate state for this component.  We observe superhump variations that are characteristic of an elliptical, precessing accretion disc. Our determination of the superhump period excess is in agreement with the established relationship between this parameter and the mass ratio, and is the most precise calibration of this relationship at low $q$. We also observe a quasi-periodic oscillation in the 2006 data, and we examine the outbursting behaviour of the system over a $4.5$ year period.
\end{abstract}

%%%%%%%%%%%%%%%%%%%%%%%%%% Begin Section 1 %%%%%%%%%%%%%%%%%%%%%%%%%%%%%
\section{INTRODUCTION}  

The AM Canum Venaticorum (AM CVn) stars are ultra-compact binaries with periods from $5.4$ \citep{Roelofs10} to $65$ minutes and optical spectra dominated by helium (see, e.g. \citealt{Nelemans05,Ramsay07} for recent reviews). These systems consist of a white dwarf accreting matter via a helium accretion disc from a significantly less massive and hydrogen deficient donor star. In order to fit within the Roche lobe it is necessary for this donor to also be at least partially degenerate. AM CVn stars offer new insights into the formation and evolution of binary star systems, with the short periods implying at least one common envelope phase in the history of the binary, and the chemical composition suggesting helium white dwarfs or CVs with evolved secondaries as possible progenitors (\citealt{Nelemans01,Nelemans10}, see also \citealt{Marsh10,Kulkarni10} for the recent identification of a possible AM CVn progenitor). Close double-degenerate binaries are also one of the proposed progenitor populations of Type Ia supernovae \citep{Tutukov81, Webbink84, Iben84} and subluminous events \citep{Perets10}. Finally, the mass transfer in these systems is thought to be driven by angular momentum loss as a result of gravitational radiation. Due to their very short periods they are predicted to be among the strongest gravitational wave sources in the sky \citep{Nelemans04}, and are the only class of binary with examples already known which will be detectable by the gravitational wave observatory LISA \citep{Stroeer06,Roelofs07}. 

Gravitational radiation has a huge influence on AM CVn systems, driving their evolution and determining their orbital period distribution, luminosities and numbers. Degenerate stars expand upon mass loss and so stable mass transfer via Roche lobe overflow causes an evolution towards longer periods. The combination of decreasing donor mass and lengthening
orbital period leads to a rapid decrease in the magnitude of the gravitational wave losses over time. There is therefore a significant drop in the mass transfer rate over the observed period range of the AM CVn population \citep{Nelemans01}. If the donor stars in AM CVn stars were completely degenerate then their masses would be a unique function of orbital period, and their mass transfer rates a function of the accretor mass and orbital period. However, the three current paradigms for the binary formation path (white dwarf mergers, \citealt{Nelemans01}; ex-helium stars, \citealt{Iben91}; CVs with evolved donors, \citealt{Podsi03}) all imply partial degeneracy, to different degrees. A partially degenerate star must be more massive than a degenerate star to fit within a Roche lobe at a given orbital period. A less degenerate donor therefore implies higher gravitational wave losses and a higher mass transfer rate. A test of the degeneracy of the donor star requires accurate mass determinations which have proved elusive, although some constraints were obtained for five systems using parallax measurements obtained with \hst \ \citep{Roelofs07b}.

The prototype AM CVn system was discovered $40$ years ago \citep{Smak67,Paczynski67}, but to date only $\sim$$25$ further objects of this class have been discovered (see, e.g., \citealt{Roelofs05, Anderson05, Anderson08, Roelofs09, Rau10}). One of these was the eclipsing system SDSS~J0926+3624 (\citealt{Anderson05}, SDSS~0926 hereafter). SDSS~0926 is currently the only eclipsing AM CVn known, and has a period of $28$ min, with eclipses lasting $\sim$$1$ min. The mean $g$-band magnitude of this system is $\sim$$19.3$ \citep{Anderson05}, but there is considerable out-of-eclipse variation, characteristic of the superhumping behaviour seen in many AM CVns and CVs which is attributed to the precession of an elliptical accretion disc \citep{Whitehurst88,Lubow91,Simpson98}.

 In $2006$ and $2009$ we took high time resolution observations of SDSS~0926 with the fast CCD camera ULTRACAM. The aim of these observations was to determine precise system parameters for this system, using techniques we have in the past successfully applied to normal CVs (e.g., \citealt{Feline04,Littlefair08,Pyrzas09,Southworth09,Copperwheat09}). Precise masses enable us to determine the degree of degeneracy of the donor star, and eclipse timings can be used to determine the angular momentum losses. We present these photometric observations in this paper, as well as additional data collected with the Liverpool Telescope, \xmm \ and the Catalina Sky Survey.

\section{OBSERVATIONS}
\label{sec:obs}

\begin{table*} 
\caption{Log of the observations.}
\label{tab:obs} 
\begin{tabular}{lllllll} 
Night           &\multicolumn{2}{c}{UT}      &Exposure		                &                           &Number of      &\\
      	        &start	    	&end        &time (s)                       &Binning	                &orbits		    &Comments\\
\hline
\multicolumn{7}{c}{\it{WHT/ULTRACAM}} \\
\\
1st Mar 2006    &$22:28$          &$04:48$  &$3$ -- $4$                     &$2 \times 2$               &$13$           &Seeing $1$ -- $2$'', some patches of cloud\\
2nd Mar         &$20:04$          &$04:49$  &$2$ -- $3$                     &$1 \times 1$/$2 \times 2$  &$18$           &Seeing $\sim 0.8$'', clear.\\
3rd Mar         &$19:56$          &$03:59$  &$3$                            &$2 \times 2$               &$16$           &Seeing $0.8$ -- $1.2$''. High humidity\\
5th Mar         &$22:50$          &$23:52$  &$3$                            &$2 \times 2$               &$2$            &Clear, but variable seeing up to $2.0$''\\
1st Jan 2009    &$00:41$          &$03:06$  &$1.8$ ($g'$,$r'$); $3.6$ ($u'$)   &$1 \times 1$               &$5$            &Data gaps due to hardware problem. \\
		&		  &	    &				    &				&		&Seeing $0.8$'' with some cloud\\
2nd Jan         &$22:52$          &$07:09$  &$1.8$ ($g'$,$r'$); $3.6$ ($u'$)   &$1 \times 1$               &$16$           &Initial poor seeing ($1.0$ -- $2.0$'') improves to $0.8$''. \\
		&		  &	    &				    &				&		&Fair transparency.\\ 
3rd Jan         &$02:57$          &$04:50$  &$1.8$ ($g'$,$r'$); $3.6$ ($u'$)   &$1 \times 1$               &$3$            &Seeing $0.8$'', Good transparency.\\
\hline
\multicolumn{7}{c}{\it{LT/RISE}} \\
\\
17 Feb 2009	&$23:08$	&$00:08$    &$30$			    &$2 \times 2$               &$2$           &Seeing $0.5$'', high humidity\\
14 Mar 		&$20:53$	&$21:53$    &$30$			    &$2 \times 2$               &$2$           &Seeing $2$ -- $3$'', photometric\\
21 Mar		&$23:09$	&$00:09$    &$30$			    &$2 \times 2$               &$2$           &Seeing $2$ -- $3$'', photometric\\
30 Mar		&$00:07$	&$02:07$    &$30$			    &$2 \times 2$               &$4$           &Seeing $2$'', photometric\\
31 Mar		&$00:56$	&$01:56$    &$30$			    &$2 \times 2$               &$2$           &Seeing $0.5$'', photometric\\
12 Apr		&$21:09$	&$22:09$    &$30$			    &$2 \times 2$               &$2$           &Seeing $0.5$'', some cloud\\
19 Apr		&$21:52$	&$22:52$    &$30$			    &$2 \times 2$               &$2$           &Seeing $2$ -- $2.5$'', photometric\\
13 May		&$21:25$	&$22:25$    &$30$			    &$2 \times 2$               &$2$           &Seeing $2$'', photometric\\
\hline
\multicolumn{7}{c}{\xmm} \\
\\
23 Nov 2006	&$10:53$ 	&$20:50$    &$35.7 \times 10^3$		    &				&$21$		&EPIC MOS \& RGS detectors\\
		&$11:15$ 	&$20:50$    &$34.0 \times 10^3$		    &				&$20$		&EPIC pn detector\\
		&$10:57$ 	&$15:37$    &$10$		    	    &				&$10$		&Optical Monitor, fast mode with UVW1 filter\\
		&$15:37$ 	&$20:51$    &$10$		    	    &				&$11$		&Optical Monitor, fast mode with UVM2 filter\\
\hline
\end{tabular}
\end{table*}

\subsection{WHT/ULTRACAM}

In $2006$ and $2009$ observations of SDSS 0926 were made with the high speed CCD camera ULTRACAM \citep{Dhillon07} mounted on the $4.2$m William Herschel Telescope (WHT). The $2006$ observations were mainly taken over a three day period in the beginning of March. Weather conditions were reasonable, with seeing $\sim$$1"$ and good transparency. The $2009$ observations were taken over three nights in January, and conditions for these winter observations were on the whole poorer, with variable seeing and transparency. Due to conditions, only a small number of orbital cycles were observed on two of the three nights. ULTRACAM is a triple beam camera and all observations were made using the SDSS $u'$, $g'$ and $r'$ filters. Average exposure times were $\sim$$3$s in $2006$ and $1.8$s in $2009$. The longer exposure time was necessary for the $2006$ observations due to the relatively low S/N of the $u'$-band data. Our $2009$ observations took advantage of a new feature in the ULTRACAM software, in which multiple $u'$-band exposures can be coadded on the CCD before readout. Two coadds were used for the majority of the data, giving a $u'$-band exposure time of $3.6$s, although this was increased during poor conditions.  The dead time between exposures for ULTRACAM is $\sim$$25$ms. The CCD was windowed in order to achieve this exposure time. A $2 \times 2$ binning was used in most of the $2006$ observations to compensate for conditions. A complete log of the observations is given in Table \ref{tab:obs}.

All of these data were reduced with aperture photometry using the ULTRACAM pipeline software, with debiassing, flatfielding and sky background subtraction performed in the standard way. The source flux was determined using a variable aperture (whereby the radius of the aperture is scaled according to the FWHM). Variations in transparency were accounted for by dividing the source light curve by the light curve of a nearby comparision star. The stability of this comparison star was checked against other stars in the field, and no variations were seen. We determined atmospheric absorption coefficients in the $u'$, $g'$ and $r'$ bands and subsequently determined the absolute flux of our targets using observations of standard stars (from \citealt{Smith02}) taken in evening twilight. We used this calibration for our determination of the apparent magnitudes of the source, although we present all light curves in flux units determined using the conversion given in \citet{Smith02}. Using our absorption coefficients, we extrapolate all fluxes to an airmass of $0$. For all data we convert the MJD (UTC) times to the barycentric dynamical timescale, correcting for light travel times.

\subsection{LT/RISE}

We supplemented our WHT/ULTRACAM data with additional observations taken through the first half of 2009 with the high speed RISE camera \citep{Steele08} mounted on the Liverpool Telescope. These observations were taken with a $2 \times 2$ binning and the RISE $V+R$ filter. Each observation was $1$h in length (except the observation on $30$ March, which was twice as long) with exposure times of $30$s. The purpose of these observations was to characterise the superhump, since this exposure time is too long to adequately sample the eclipse. We reduced these data using aperture photometry as with the WHT/ULTRACAM data, using the ULTRACAM pipeline software.

\subsection{XMM-NEWTON}

SDSS~0926 was observed with \xmm \ on $23$ November $2006$. It was observed for $34.0$ ksec in the EPIC pn detector and $35.7$ ksec in the EPIC MOS detectors. It was detected with a mean count rate of 0.033$\pm$0.001 ct/s in the EPIC pn and 0.021$\pm$0.008 ct/s in the EPIC MOS (1+2), but was too faint in X-rays to be detected in the RGS detectors. The particle/X-ray background was low during the course of the observation. The Optical Monitor (OM) was configured in fast-mode and the observation time was split between the UVW1 (2450--3200\AA) and UVM2 (2050--2450\AA) filters. The source was detected with a mean count rate of 0.68 ct/s in UVW1 and 0.24 ct/s in UVM2.

The X-ray data were processed using the {\sl XMM-Newton} {\sl Science Analysis Software} (SAS) v9.0. For the EPIC detectors, data were extracted using an aperture of 30$^{''}$ centered on the source
position. Background data were extracted from a source free region. The background data were scaled and subtracted from the source data. The OM data were reduced using {\tt omfchain}.

\subsection{The Catalina Real-Time Transient Survey}

The Catalina Sky Survey \citep{Larson98} is a search for near-Earth objects using the $0.7$m $f/1.9$ Catalina Schmidt Telescope north of Tucson, Arizona. This survey uses a single unfiltered $4$k $\times$ $4$k CCD with $2.5$'' pixels, giving an $8$ deg$^2$ field of view. The Catalina Real-Time Transient Survey (CRTS; \citealt{Drake09}) began analysing these data in real-time in November $2007$ for optical transients. 

The CRTS dataset contains $202$ separate observations of SDSS~0926 between $10$ November $2004$ and $11$ June $2010$. Each observation is $30$s in length, and they are divided up into groups of (typically) $4$ observations taken over a $\sim$$30$ min period. These data were reduced following \citet{Drake09}.

\section{RESULTS}
\label{sec:resultsa}

\subsection{WHT/ULTRACAM light curves}
\label{sec:lightcurves}

\begin{figure*}
\centering
\includegraphics[angle=270,width=1.0\textwidth]{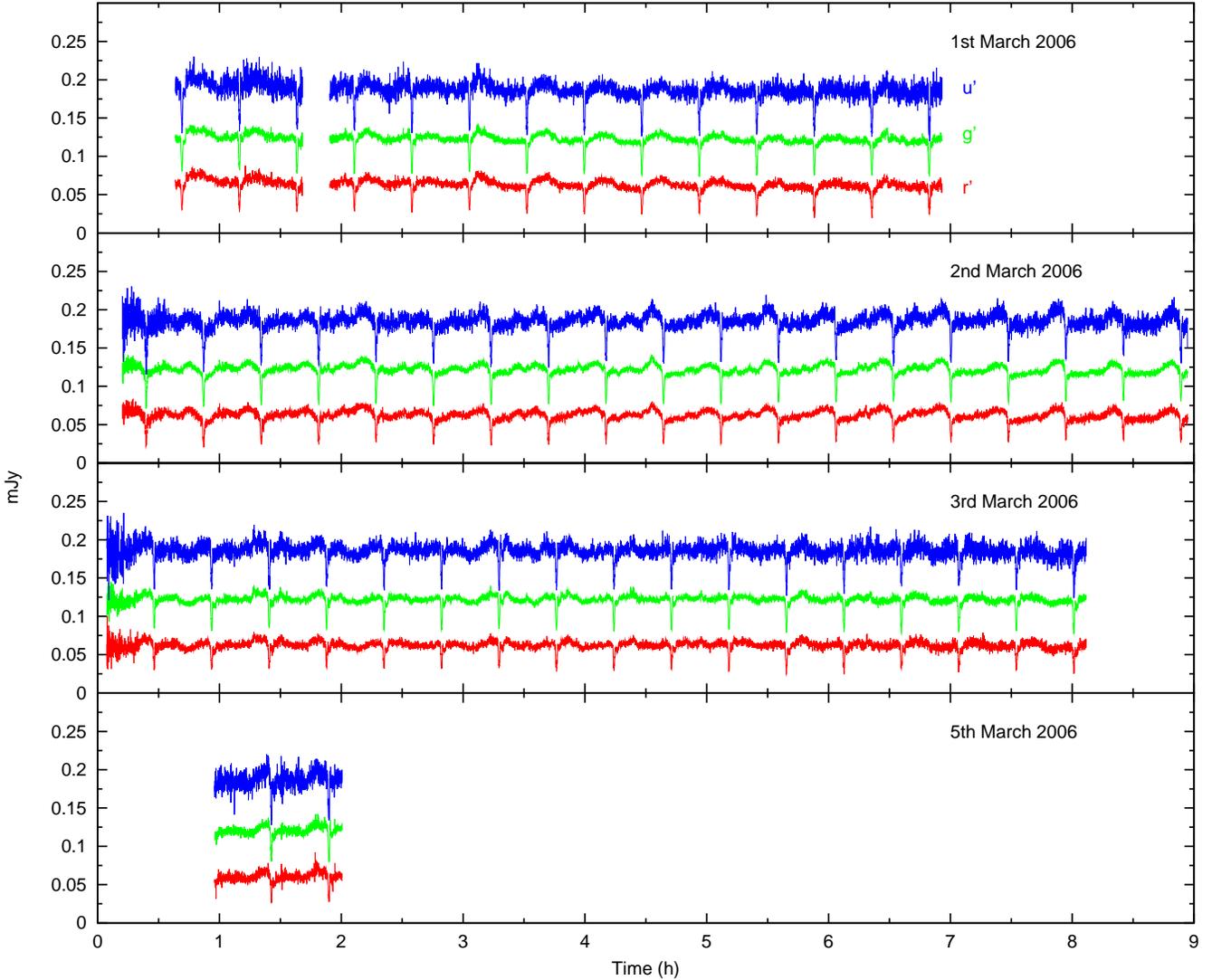}
\hfill
\caption{Light curves of SDSS0926, observed in March 2006 with WHT/ULTRACAM. All data were collected simultaneously in the $u'$- (top, blue), $g'$- (middle, green) and $r'$-bands (bottom, red). For clarity we apply offsets of $0.05$mJy to the $g'$-band data and $0.1$mJy to the $u'$-band data. The gaps in the first night of data are due to poor weather conditions. } \label{fig:lightcurves_06} \end{figure*}

\begin{figure*}
\centering
\includegraphics[angle=270,width=1.0\textwidth]{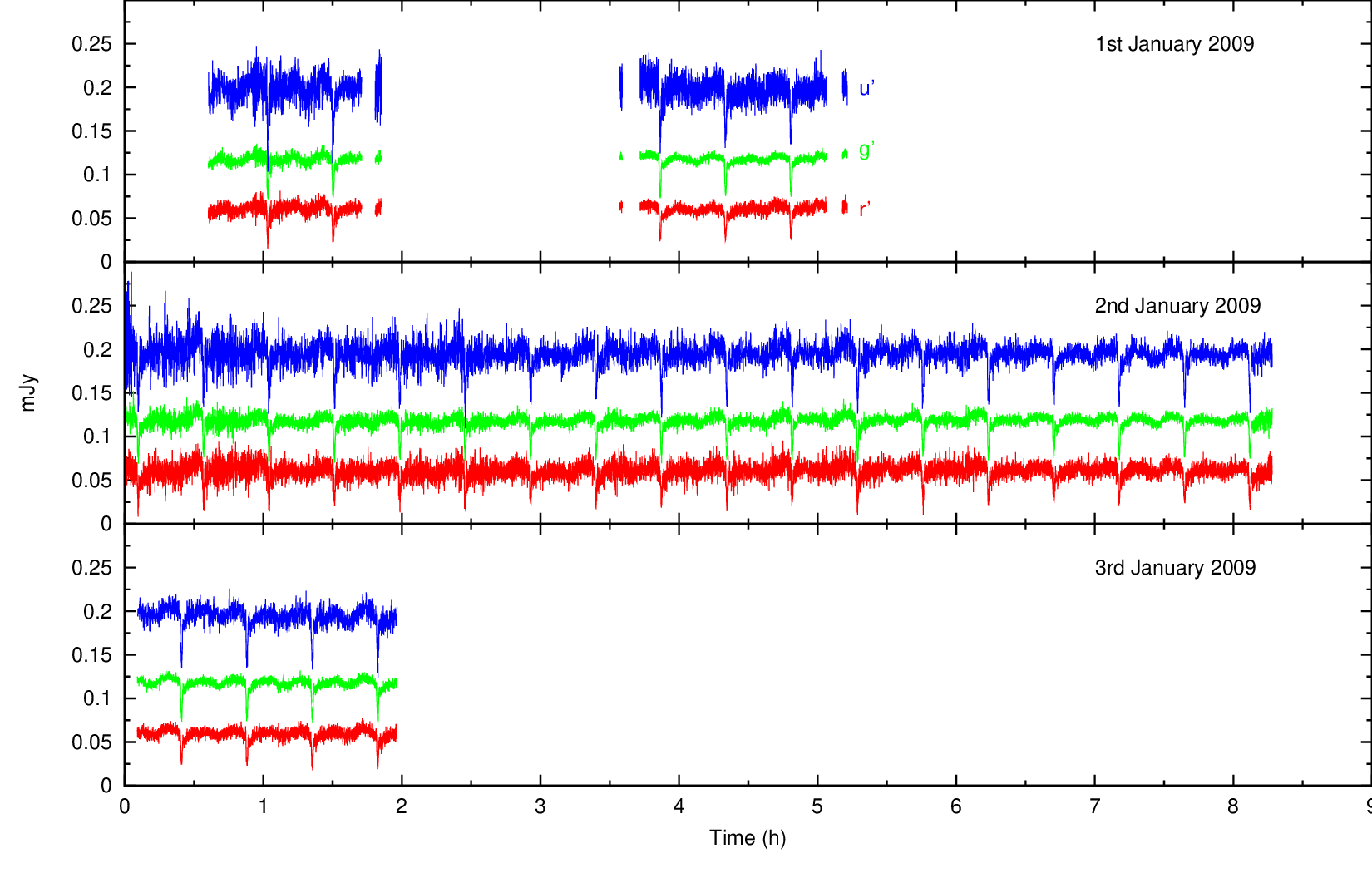}
\hfill
\caption{Light curves of SDSS0926, observed in January 2009 with WHT/ULTRACAM. All data were collected simultaneously in the $u'$- (top, blue), $g'$- (middle, green) and $r'$-bands (bottom, red). For clarity we apply offsets of $0.05$mJy to the $g'$-band data and $0.1$mJy to the $u'$-band data. The gaps in the first night of data are due to poor weather conditions and a hardware fault.} \label{fig:lightcurves_09} \end{figure*}

\begin{figure*}
\centering
\includegraphics[angle=270,width=1.0\textwidth]{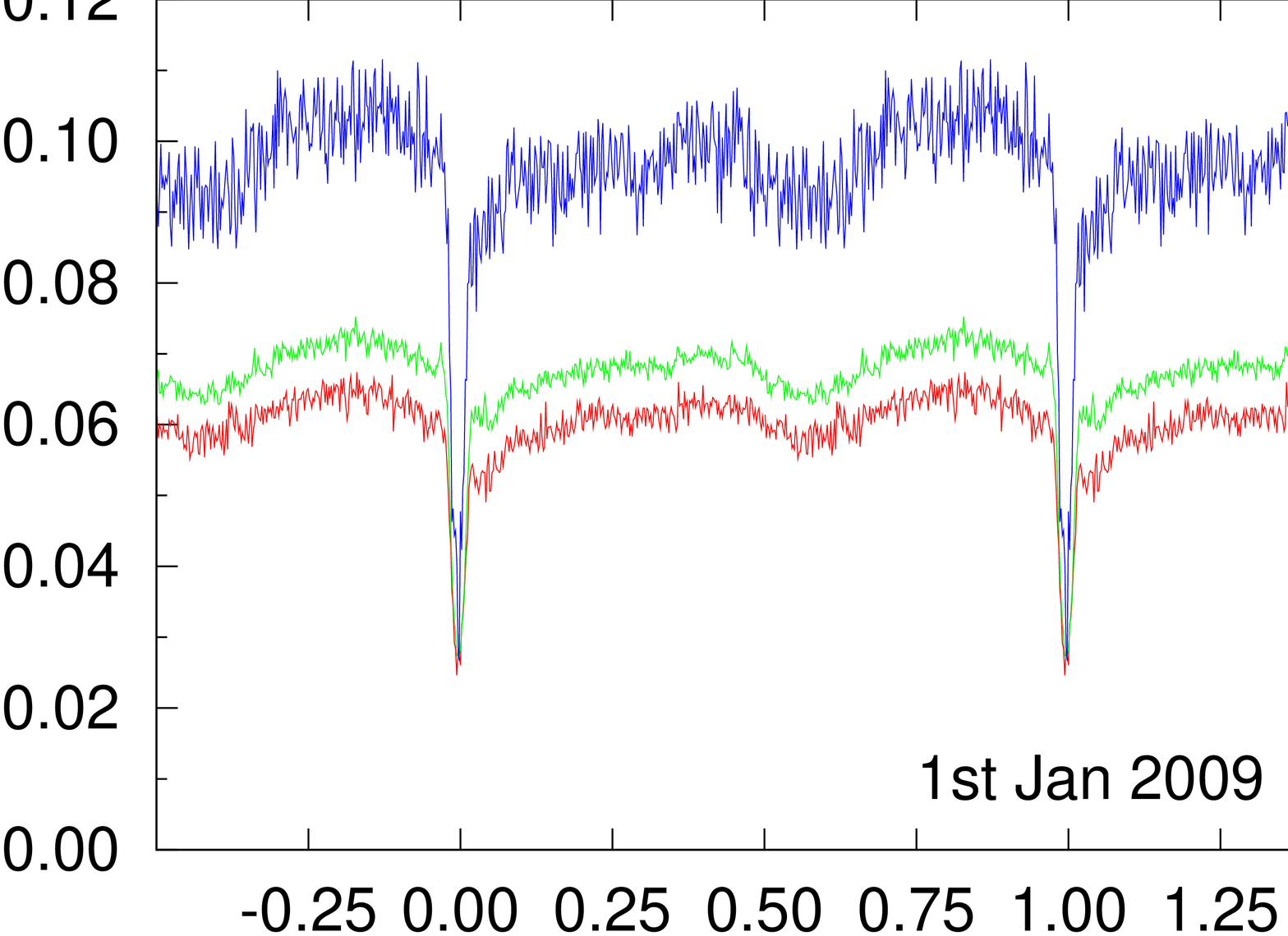}
\hfill
\vspace{5mm}
\caption{Phase folded and binned light curves, showing the superhump variation from night-to-night. In the top row we plot the first three nights of data collected in 2006. In the bottom row we plot the three nights of data taken in 2009. We plot separately the data in the $u'$- (top, blue), $g'$- (middle, green) and $r'$-bands (bottom, red).} \label{fig:superhump} \end{figure*}

The March 2006 data are plotted in Figure \ref{fig:lightcurves_06} and the January 2009 data are plotted in Figure \ref{fig:lightcurves_09}. Additionally, we phase-folded the data on a night-by-night basis using the ephemeris given in Section \ref{sec:ephemeris}, and plot the results in Figure \ref{fig:superhump}. We omit from this plot the short section of data collected on $5$th March $2006$.

If we first examine the $2006$ data, it is apparent that there are gross differences in the light curve from night to night. The shape of the eclipse features remains the same, but the superhump precesses through the light curve, and so we see the peak of the superhump emission at different phases on different nights. On $1$ March the peak of the superhump is soon after the eclipse. On $2$ March it is shortly before the eclipse, and on 3 March it is not immediately apparent, but the shape of the light curve before and after the eclipse suggests that the superhump and the eclipse are approximately superimposed. If we examine the eclipse feature itself, we see that the primary eclipse is immediately followed by a distinct second, smaller eclipse (this is most apparent in the $3$ March data). We will show in Section \ref{sec:phasefold} that these two eclipses are of the white dwarf and the bright spot, respectively. The eclipses are preceded by a small orbital `hump' caused by the bright spot moving into the field of view. This is not immediately apparent since the bright spot is relatively weak in these data, so the out-of-eclipse variation is dominated by the superhump.
As well as the superhump and eclipse features we see the stochastic `flickering' variation that is characteristic of accreting sources. This variation is mitigated to some degree in the phase-folded lightcurves (Figure \ref{fig:superhump}). Following \citet{Smith02} we find the mean magnitudes outside of eclipse to be $19.05 \pm 0.10$, $19.24 \pm 0.07$ and $19.39 \pm 0.08$ in $u'$, $g'$ and $r'$ respectively.

In contrast to $2006$, in the $2009$ data the shape of the out-of-eclipse light curve is roughly constant from night to night: 
we do not see the large variations caused by a superhump component precessing through the light curve. The shape of the light curve on all three nights is most similar to the $2$ March 2006 data, with the peak of the emission shortly before the eclipse. Since the position of this peak does not vary from night-to-night it is most likely due to the bright spot, and thus there seems to be no significant superhump contribution in these data. The mean magnitudes outside of eclipse are $18.94 \pm 0.13$, $19.31 \pm 0.07$ and $19.43 \pm 0.11$ in $u'$, $g'$ and $r'$ respectively, consistent with the $2006$ values. Note also that the out-of-eclipse variation is double-humped, with a peak at a phase of $\sim$$0.3$ as well as the main peak at $\sim$$0.8$. The most likely explanation for this is that bright spot is vertically extended, or disc is optically thin, so we are seeing emission from the bright spot all the way round the orbit.

\subsection{Non-orbital variability in the ULTRACAM data}
\label{sec:nonorbvar}

In Section \ref{sec:analysis} we use the high time resolution ULTRACAM data to make precise parameter determinations. However, first it is necessary to examine the non-orbital variability in this system. We begin by examining quasi-periodic variability in the ULTRACAM data. Secondly, it is important to characterise the superhumps present in the 2006 observations, so these features can be subtracted from the lightcurves. 

%The superhumps are a transient phenomenon that is driven by outbursts, so finally we use the LT and CRTS data to examine the outbursting history of this system.

\subsubsection{Quasi-coherent variability}
\label{sec:qpo}

\begin{figure}
\centering
\includegraphics[angle=270,width=1.0\columnwidth]{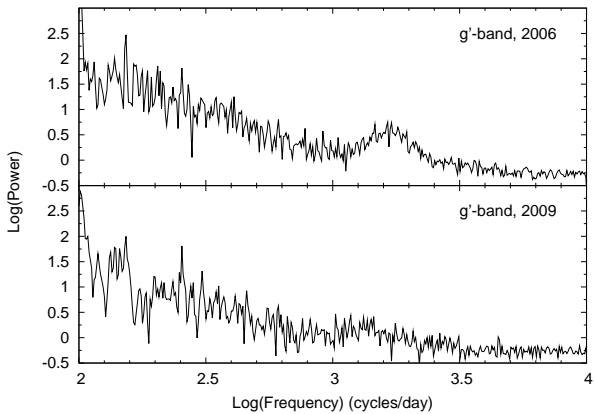}
\hfill
\caption{Lomb-Scargle periodgrams for the $g'$-band 2006 (top) and 2009 (bottom) WHT/ULTRACAM datasets. We convert both the y- and x-axes to a logarithm scale, and then uniformly bin the data along the x-axis (frequency). In the $2006$ dataset we see a quasi-periodic oscillation (QPO) at a frequency of $\sim$$1700$ \ cycles/day (P=$\sim$$50$s).} \label{fig:qpo} \end{figure}

In Figure \ref{fig:qpo} we plot Lomb-Scargle periodograms \citep{Press02} for the complete 2006 and 2009 $g'$-band datasets obtained with WHT/ULTRACAM. In the 2006 data we detect a signal at a frequency of $\sim$$1700$cycles/day (P$\sim$$50$s), although it is incoherent and spread over a wide frequency range. We estimate the quality factor (the peak centroid frequency divided by its full width at half maximum) of this signal to be $Q \sim$$4$ in the $g'$-band data. We computed the periodograms for each of the first three 2006 nights separately and we detected this signal every night. The signal is high in the $g'$-band, and is barely detected in the $u'$ band. There is possibly a signal at the lower frequency of $\sim$$1400$ \ cycles/day in the $2009$ data, but it is much weaker than the $2006$ signal. We did not find any signals in the periodograms at higher frequencies beyond the $5000$ cycles/day range plotted in Figure \ref{fig:qpo}.

Similar quasi-coherent variability was first observed in CVs some decades ago \citep{Warner72, Patterson77}, and has since been observed in many CVs and X-ray binaries (see \citealt{Warner08} for a recent review). The peak period of the signal we detect in the 2006 data is low for a QPO, but this may be due to the fact that the accretion disc in an ultracompact binary such as SDSS~0926 is much smaller and less massive than the disc in conventional CV systems.

\subsubsection{Superhumps}
\label{sec:superhump}
 
\begin{table} 
\caption{Superhump parameters. We list the period and amplitude of the primary sine function fitted to the complete 2006 dataset, for each of the three bands. We list also the period excess $\epsilon = (P_{SH} - P_{Orb}) / P_{Orb}$, where $P_{SH}$ and $P_{Orb}$ are the superhump and orbital periods, respectively.}
\label{tab:wdparams} 
\begin{tabular}{ll@{\,$\pm$\,}ll@{\,$\pm$\,}ll@{\,$\pm$\,}l} 
Filter  &\multicolumn{2}{c}{Period}  &\multicolumn{2}{c}{Amplitude}  &\multicolumn{2}{c}{Period excess ($\epsilon$)}\\
        &\multicolumn{2}{c}{(min)}   &\multicolumn{2}{c}{(mJy)}\\
\hline
$u'$     &$28.560$ &$0.003$     &$44.1$ &$0.8$   &$0.00875$ &$0.00010$ \\    
$g'$     &$28.558$ &$0.001$     &$50.6$ &$0.5$   &$0.00869$ &$0.00005$ \\
$r'$     &$28.553$ &$0.001$     &$60.1$ &$0.5$   &$0.00849$ &$0.00005$ \\
\hline		
\end{tabular}
\end{table}

\begin{figure}
\centering
\includegraphics[angle=270,width=0.9\columnwidth]{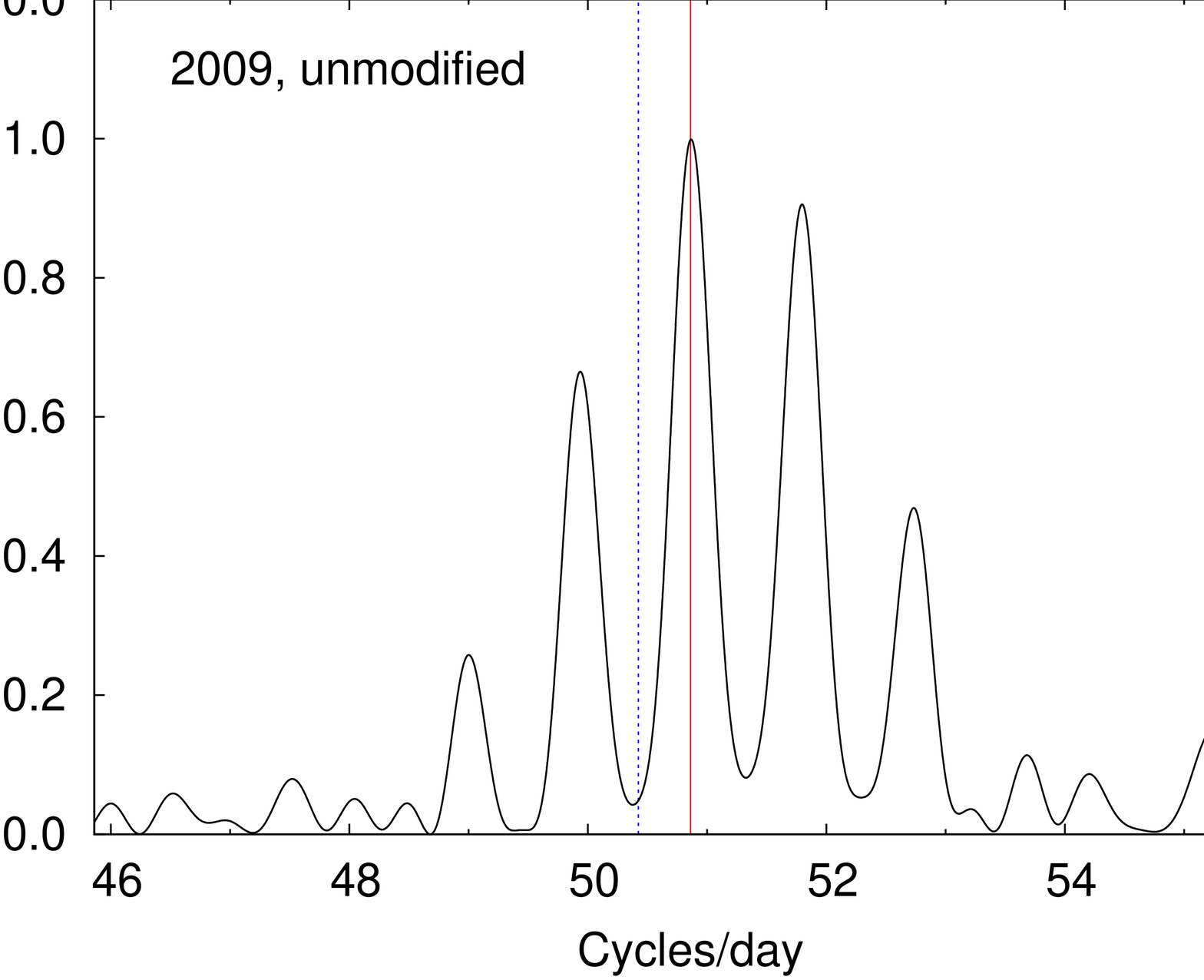}
\vspace{5mm}
\caption{Lomb-Scargle periodograms for the $g'$ band data. We plot the power around the orbital frequency, with the peak of each distribution normalised. The vertical lines mark the frequency of the superhump (blue, dotted) and the orbital (red, solid) modulations. The four periodograms are (from top to bottom): (i) The $2006$ data, showing both the superhump and orbital modulations. (ii) The $2006$ data with the eclipse and bright spot features masked. (iii) The $2009$ data with the superhump subtracted, using the method described in Section \ref{sec:superhump}. (iv) The $2009$ data.} \label{fig:powerspec} \end{figure} 

Figure \ref{fig:powerspec} shows $g'$-band Lomb-Scargle periodgrams in the vicinity of the orbital frequency, with various manipulations applied. The top panel shows the combined $2006$ dataset, with no modification. The second panel uses the same dataset, but with the eclipse features masked. A phase range of $0.18$ is masked, centred on the mid-point of the white dwarf eclipse. This phase range is sufficient to cover the eclipses of both the white dwarf and bright spot and the peak of the bright spot emission. The third panel shows the same dataset with the eclipse features unmasked but with the superhump subtracted (see below). The bottom panel shows the combined $2009$ dataset, with no modification. In all four panels, vertical lines mark the positions of the superhump and orbital frequencies.

If the top panel of Figure \ref{fig:powerspec} is examined, it can be seen that the power due to superhumps is clearly apparent, stronger than the orbital signal and peaking at a slightly lower frequency. The two signals are confused in this first panel, but the superhump is seen as being clearly distinct in the second panel, in which the majority of the orbital modulation is masked.

In order to determine the parameters of the $2006$ superhump, we fitted a model to the combined dataset with the eclipse features masked. The model consists of a combination of six 4-parameter sine functions: three for the superhump, fitting the primary frequency and the second and third harmonics, and three for harmonics of the orbital period, so as to fit any residual signal left after the masking of the eclipse features. In the third panel of Figure \ref{fig:powerspec} the periodogram for the unmasked dataset with the superhump components subtracted is plotted. We see that our model fits do a good job of cleaning the superhump signal from the data.

Our findings for the superhump period are listed in Table \ref{tab:wdparams}.  The uncertainties on these periods were determined from fits to a series of sample datasets derived from the originals using the bootstrap method \citep{Efron79,Efron93}. The amplitudes of the superhump harmonics are $< 10\%$ of the amplitude values for the primary frequency. We find the period of the variation to be consistent at the $1 \sigma$ level for the $u'$- and $g'$-bands, but the $r'$-band period is lower. This inconsistency is probably due to some extra intrinsic variability, such as accretion-driven flickering. The amplitude of the modulation increases at longer wavelengths. We list also in this table the period excesses $\epsilon$ in each band, using the orbital period given in Section \ref{sec:ephemeris}. 

Finally, the fourth panel of Figure \ref{fig:powerspec}  shows the $2009$ data. This plot shows a clean signal at the orbital frequency, confirming that there is no superhump modulation in these data. Following these observations, we obtained a series of $1$h light curves using LT/RISE over the first half of $2009$. The purpose of these observations was to examine the long term behaviour of the superhump, although with hindsight they were perhaps too short. Two of these light curves were in the immediate aftermath of an outburst in this system, and we will discuss these separately in Section \ref{sec:outburst}. Of the remaining light curves, few show clear evidence for the superhump. On $17$ February and $21$ March there is a `hump' just before the eclipse, but this could be due to the bright spot. Most of the remaining quiescent light curves show little out-of-eclipse variation. One exception is the light curve obtained on $19$ April, which shows a clear brightening immediately following the eclipse, which can only be explained by the superhump. This light curve was obtained $21$ days after the detection of the outburst in this system. There was no clear evidence for the superhump obtained seven days prior to this one, on $12$ April.

\begin{figure}
\centering
\includegraphics[angle=270,width=1.0\columnwidth]{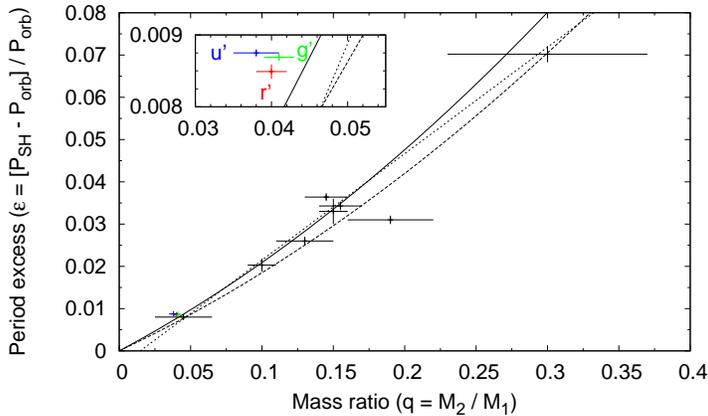}
\hfill
\caption{Mass ratio $q$ versus the superhump period excess $\epsilon$. The blue, green and red points show our $u'$-, $g'$- and $r'$-band determinations, respectively. The region around these points is magnified in the inset. The solid line is the $\epsilon = 0.18 q + 0.29 q^2$ relationship proposed by \citet{Patterson05}. The dashed line is the $\epsilon = 0.16q + 0.25q^2$ relation proposed by \citet{Kato09}. The dotted line is the linear $q(\epsilon) = (0.114 \pm 0.005) + (3.97 \pm 0.41) \times (\epsilon - 0.025)$ relation proposed by \citet{Knigge06}. The black points are the eclipsing CVs listed as calibration sources in Table 7 of \citet{Patterson05}.} \label{fig:patterson} \end{figure}

\citet{Patterson05} suggested  $\epsilon = 0.18 q + 0.29 q^2$ as an empirical relationship between the superhump period excess $\epsilon = (P_{SH} - P_{Orb}) / P_{Orb}$, and the mass ratio $q$. This relationship was calibrated using measurements of a series of eclipsing systems, listed in Table 7 of \citet{Patterson05}. The relation is pinned by setting $\epsilon = 0$ when $q = 0$, but this is an assumption and not empirically determined: of the calibration systems only KV UMa has a mass ratio $< 0.05$ and this determination is very uncertain, and so the calibration is potentially poor at low mass ratios. SDSS~0926 is therefore potentially a strong test of this relationship, although it has been argued that it would not apply to AM CVn systems \citep{Pearson07}. In Figure \ref{fig:patterson} we reproduce Figure $1$ from \citet{Patterson05}, adding our measurements from the $2006$ data in $u'$, $g'$ and $r'$, using the values given in Tables \ref{tab:wdparams} and \ref{tab:params}. As well as the Patterson relation, we also plot the slightly modified relation proposed by \citet{Kato09} ($\epsilon = 0.16(2)q + 0.25(7)q^2$) and the linear relation proposed by \citet{Knigge06} ($q(\epsilon) = (0.114 \pm 0.005) + (3.97 \pm 0.41) \times (\epsilon - 0.025)$). The Knigge relation does not assume $\epsilon = 0$ when $q = 0$. Our measurements of the period excess in SDSS~0926 are consistent with all of these relations to within their uncertainties, and are closest to the \citet{Patterson05} relation, which suggests the assumption of $\epsilon = 0$ when $q = 0$ is reasonable. 

\subsection{Outbursting behaviour}
\label{sec:outbursts}

The superhumps are a transient phenomenon that is driven by outbursts, so here we examine the outbursting history of this system, presenting the first observations of SDSS~0926 in outburst. We begin by discussing observations taken with the LT in the immediate aftermath of an outburst in March 2009. We go on to examine the long-term outbursting behaviour using five years of CRTS observations.

\subsubsection{The March 2009 outburst}
\label{sec:outburst}

\begin{figure*}
\centering
\includegraphics[angle=270,width=0.8\textwidth]{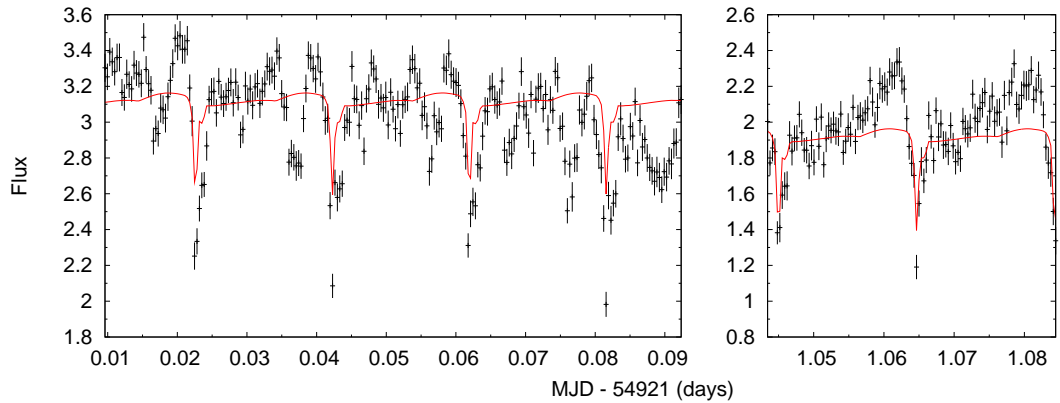}
\hfill
\caption{LT/RISE light curves obtained on 30 and 31 March 2009: one day and two days after an outburst in this source was first detected, respectively. We plot these data in flux units, with the mean quiescent out-of-eclipse flux set to unity. The solid red line is a model fit to our quiescent, superhump-subtracted LT/RISE observations with an arbitrary flux offset, which we plot for comparative purposes.} \label{fig:outburst} \end{figure*}

On $29$ March 2009 it was discovered that SDSS~0926 was in outburst. We obtained $2$h of data with LT/RISE on the subsequent night, and a further $1$h on the night after that. We plot these data in Figure \ref{fig:outburst}. The flux scale for these data is such that the mean, quiescent out-of eclipse flux is equal to $1$. We also plot the model fit (described in Section \ref{sec:phasefold}, below) to our quiescent superhump subtracted LT/RISE observations for comparison. We apply an arbitrary flux offset to this model light curve so as to overlay it on the outburst data. We see that at the beginning of our $30$ March observation the system was a factor of $\sim$$3.5$ brighter than during quiescence and declining quickly: this has dropped to a factor of $3$ by the end of the $2$h observation. The following night this has dropped to a factor of $2$. There is still some evidence of the outburst in our next observation on $12$ April: the out-of-eclipse flux in this light curve is $\sim$$10$\% greater than the mean level.

\begin{figure}
\centering
\includegraphics[angle=270,width=1.0\columnwidth]{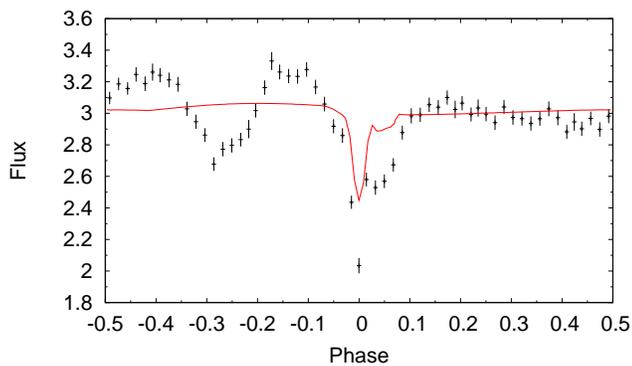}
\hfill
\caption{The LT/RISE data collected one day after the March 2009 outburst, phase-folded using the ephemeris given in Section \ref{sec:ephemeris}. We plot these data in flux units, with the mean quiescent out-of-eclipse flux set to unity. The solid red line is the quiescent, superhump-subtracted model light curve  with an arbitrary flux offset, which we plot for comparative purposes.} \label{fig:liv_ff} \end{figure}

We phase-folded the $30$ March data using the ephemeris given in Section \ref{sec:ephemeris}, and plot the results in Figure \ref{fig:liv_ff}. Again, we plot a quiescent model light curve for comparison. In this plot the structure in the light curve after the outburst is more evident. There are two main features: the primary eclipse centred on a phase of $0$, and a separate, smaller dimming in the light curve centred on a phase of $\sim$$-0.25$. If we examine the primary eclipse first we see it is much deeper and wider than the quiescent white dwarf eclipse. The eclipse width is consistent with this being an eclipse of the accretion disc. The eclipse is asymmetric however, and there is a clear `step' in the egress. This suggests an uneven distribution of flux over the surface of the accretion disc itself, or (perhaps more likely) this step could be due to the bright spot. The step height of this egress is much larger than the bright spot egress during quiescence however, and so if this feature is attributed to the spot it would imply an enhanced mass transfer during the outbursting state, or an increased viscous heating at the bright spot position.

The second unusual feature in the phase-folded light curve is the dimming centred on a phase of $\sim$$-0.25$. This dimming begins at approximately the same time that the bright spot begins to come into view in the quiescent light curve. The phase of this feature is such that the eclipsing component cannot be either the white dwarf, the donor or the bright spot. We suggest this feature may be due to a warped accretion disc \citep{Pringle96}, the distortion being induced by the outburst. The dimming we see can therefore be explained by obscuration of the white dwarf or inner disc by the accretion disc itself. This is a very short lived feature: we see no evidence for it in the data collected on the subsequent night.

Finally, as we discussed in Section \ref{sec:superhump} we detect superhumps in this system in data obtained $21$ days after the outburst. This is consistent with the expectation that, since the superhumps in this system are not permanently present, they are induced by the perturbation of the disc by the outburst.

\subsubsection{Long-term variations}

\begin{figure}
\centering
\includegraphics[angle=270,width=1.0\columnwidth]{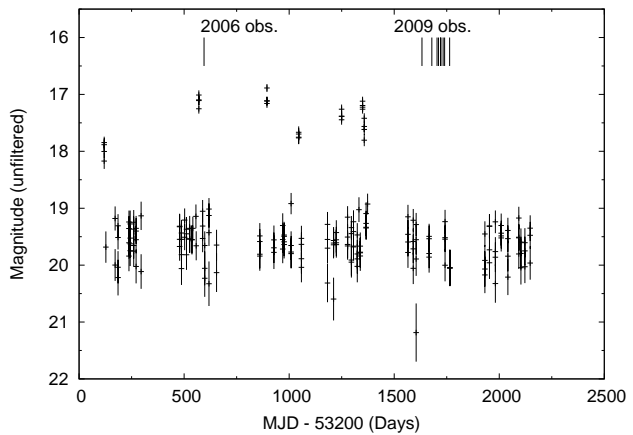}
\hfill
\caption{Observations made by the Catalina Real-Time Transient Survey (CRTS). These data were collected between $10$ November $2004$ and $11$ June $2010$, and six outbursts were observed over that period. The solid lines at the top of the plot indicate the times of the 2006 and 2009 observations with the WHT and LT.} \label{fig:css} \end{figure}

We plot in Figure \ref{fig:css} the unfiltered data collected by the Catalina Real-Time Transient Survey (CRTS). These data were collected between $10$ November $2004$ and $11$ June $2010$. We observe a total of either six or seven outbursts in these data, but the time sampling is such that they do not give an exhaustive picture of the outbursting behaviour of the source over this time period: there are no observations around the time of the March 2009 outburst, for example, and so this outburst is missed. Note that we do detect an outburst $\sim$$25$ days before the 2006 ULTRACAM observations. It is likely that this outburst induced the superhumps we see in these data by perturbing the disc.

The CRTS data is split up into blocks of four $30$s observations taken over a $\sim$$30$ min period. There is typically $> 20$ days between these blocks, and so in most cases the source has returned to quiescence by the time of the observation block subsequent to the detection of the outburst. The exception is the sixth and seventh outburst: these are separated by only eight days with no intermediate points, so may be one long outburst or two separate ones. An $8+$ day outburst is long considering the orbital period of this system: if we scale by orbital period this would be the equivalent of a $2$ month long outburst in a $P = 3.5$ h dwarf nova. However, outbursts of $\sim$$12$ days have been observed in KL Dra, a $P = 25$ min AM CVn system \citep{Ramsay10}.

The mean magnitude of the quiescent, out-of-eclipse points is $19.33 \pm 0.31$. The outbursts range in magnitude from $17.11 \pm 0.08$ to $16.81 \pm 0.12$. Note that for all six outbursts it is unclear how much time has elapsed between the initial rise and the observation, and so these measurements set a lower limit on the peak brightness. Our LT/RISE observations of the March $2009$ outburst showed a factor $\sim$$3$ enhancement in flux a day after the initial detection, and a factor $\sim$$2$ enhancement a day after that. This is relatively modest compared to the outbursts observed in the CRTS data. We conclude that the March $2009$ outburst was a relatively minor one for this source. An outburst of this size would only be detectable in the CRTS data within $1$ -- $2$ days of the initial rise. The average outburst recurrence time is therefore difficult to determine. If we exclude the two outbursts which occur within $8$ days of each other the time between observed outbursts ranges from $104$ to $449$ days, but since we may have missed a number of outbursts the actual recurrence time may be shorter. For comparison, an outburst cycle of $\sim$$60$ days was observed in KL Dra \citep{Ramsay10}.

\subsection{XMM-Newton observations}
\label{sec:xmmanalysis}

Observations have shown that the ultraviolet luminosities of AM CVns are high \citep{Ramsay05}, hence to fully understand the energy budget of these systems it is necessary for us to make observations at X-ray and ultraviolet wavelengths. We introduce here the data we obtained with \xmm, and go on to use the UV OM data to determine the temperature of the primary white dwarf.

\subsubsection{X-ray data}
\label{sec:xmmxray}

We extracted light curves from the EPIC pn and both EPIC MOS cameras and combined them into one light curve, which was folded using the ephemeris given in Section \ref{sec:ephemeris}. We found  no  evidence for an eclipse in the X-ray light curve. To investigate this further we searched for periods using a Discrete Fourier Transform and phase dispersion methods and found no clear signal of an
eclipse. 

Using the original X-ray light curve as a benchmark, we added an eclipse of given depth to the light curve. We generated 100 light curves for a given eclipse depth. We found that for a partial eclipse with a depth of $< 90$\%, we would likely not detect the eclipse. Even for a total eclipse we would have only a 70\% probability of detecting the eclipse.

We extracted integrated X-ray spectra from the EPIC pn and MOS detectors (with corresponding background spectra) and fitted them simultaneously using an absorbed thermal plasma model. Unlike other AM
CVn systems \citep{Ramsay05,Ramsay06}, we found that varying the metal abundance did not significantly improve the goodness of fit (probably since the signal-to-noise ratio of the spectra was relatively low). We obtained a good fit to the data, and the derived spectral parameters are listed in Table \ref{tab:xray}.

The best-fit value for the absorption is high (1.3$\pm0.3\times10^{21}$\pcmsq) for an object at a high galactic latitude (+46$^{\circ}$). In contrast, the absorption to the edge of the Galaxy is
$\sim1.4\times10^{20}$ \pcmsq \citep{Dickey90}. If we fix the column density parameter in the spectral fits to the \citet{Dickey90} value the resulting fit (\rchi=1.96, 104 dof) was poorer at a confidence level of $>99.99\%$. If this `extra' absorption originates in SDSS~0926 itself it may be due to viewing the boundary layer through the disc, since the binary inclination is high. Finally, we calculate the unabsorbed bolometric X-ray luminosity using the distance given in Section \ref{sec:components}, and find it to be $L_x = 6.9 \times 10^{30}$\ergss.

\begin{table}
\begin{center}
\begin{tabular}{ll}
\hline
$N_{H}$ 				& 1.3$\pm0.3\times10^{21}$ \pcmsq \\ 
$\alpha$ 				& 1.8$_{-0.5}^{+\infty}$ \\
$kT_{max}$ 				& 30 keV (fix) \\ 
Observed flux 0.1-10keV 		& 1.2$^{+2.8}_{-0.3}\times10^{-13}$ \ergscm \\ 
Unabsorbed \\ bolometric flux 		& 2.7$_{-0.8}^{+5.8}\times10^{-13}$ \ergscm \\
Unabsorbed \\ bolometric luminosity 	& 3.2$_{-0.9}^{+6.8}\times10^{29}$ \ergsd \\\\ 
					& \rchi=1.03 (103 dof)\\
\hline
\end{tabular}
\caption{The X-ray spectral fit parameters derived using a simultaneous fit to the EPIC pn and MOS spectra.}\label{tab:xray}
\end{center}
\end{table}

\subsubsection{Ultraviolet data}
\label{sec:xmmuv}

\begin{figure}
\centering
\includegraphics[width=1.0\columnwidth]{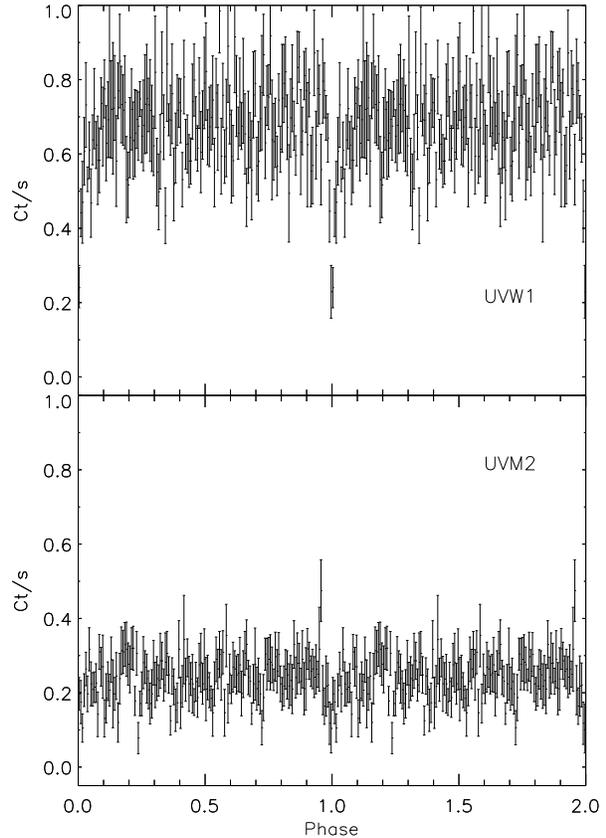}
\hfill
\caption{Phase folded UV lightcurves obtained using {\sl XMM-Newton} OM. The top and bottom plots are through the UVW1 and UVM2 filters, respectively.} 
\label{fig:uv} 
\end{figure}

We binned the OM light curves into $5$s bins and folded them on the ephemeris given in Section \ref{sec:ephemeris}. The resulting light curve is plotted in Figure \ref{fig:uv}. The eclipse is clearly seen in the
UVW1 filter, where the count rate is consistent with zero at $\phi$=0.0 in a single bin. In the shorter wavelength filter, UVM2, the signal-to-noise ratio is lower, but there is some evidence for a dip
at $\phi$=0.0, although it is not total. We searched for other periods using a Discrete Fourier Transform \citep{Deeming75} and phase dispersion methods \citep{Stellingwerf78} and found no other significant periods in the light curves.

We derived UV fluxes by converting the count rate in the two UV filters to a flux by assuming a conversion factor which was derived using observations of white dwarfs using {\sl XMM-Newton} OM.  A mean
count rate of $0.71 \pm 0.02$ ct/s in the UVW1 filter corresponds to a flux of 3.2$\times10^{-16}$ \ergscm$/$\AA, while $0.27 \pm 0.01$ ct/s in the UVM2 filter corresponds to a flux of 5.3$\times10^{-16}$ \ergscm$/$\AA. We calculate the ultraviolet luminosity following the approach of \citet{Ramsay05, Ramsay06}, by assuming a blackbody flux distribution and fixing the normalisation to give the inferred de-reddened ultraviolet flux in the two filters. This gives an ultraviolet luminosity of $L_{uv} = 1$ -- $3 \times 10^{32}$\ergss. Both this luminosity and the X-ray luminosity $L_x$, given in the previous section, are consistent with the findings for other AM CVn systems of similar period \citep{Ramsay06}.

\begin{figure}
\centering
\includegraphics[angle=270,width=1.0\columnwidth]{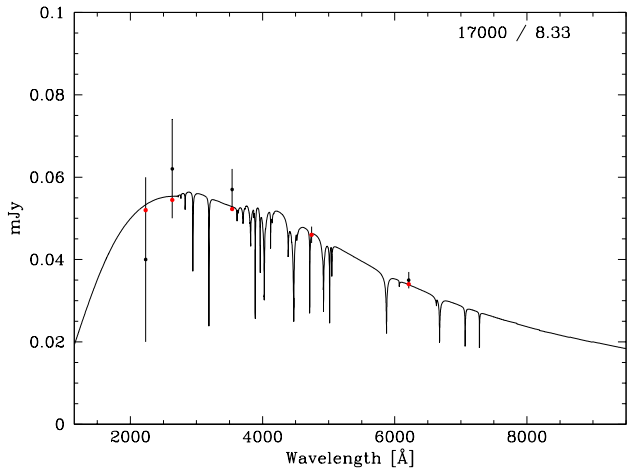}
\hfill
\caption{We determine the white dwarf temperature by fitting our measurements of the white dwarf flux in the optical and UV to a library of synthetic spectra \citep{Gaensicke95}. The spectrum we plot here is for a $17,000$K white dwarf, and is the best fit to our flux determinations. The numbers in the top right corner are the temperature and $\log g$ for the model fit.} \label{fig:wdtemp} \end{figure}

We used the UV and optical fluxes to determine the effective temperature of the white dwarf, by estimating the contribution of the white dwarf in each band. In the three ULTRACAM bands this parameter is determined through our model fits, as discussed in Section \ref{sec:analysis}. In the UV data we estimated the white dwarf contribution from the eclipse depth. We made a correction to compensate for the fact that the eclipse is only partial, by determining the fraction of the white dwarf's surface area that is obscured during the eclipse in our optical fits. This correction is small though, as evidenced by the fact that the UVW1 flux reaches zero during the eclipse, within the errors. We calculated the white dwarf contributions to be UVM2 $\sim$$0.04 \pm 0.02$, UVW1 $\sim$$0.062 \pm 0.012$, $u' =  0.057 \pm 0.005$, $g' = 0.046 \pm 0.002$ and $r' = 0.035 \pm 0.002$ mJy. We then fitted these fluxes with the white dwarf model atmospheres introduced in \citet{Gaensicke95}. We find a white dwarf temperature of $17,000$K to be consistent with our measurements (Figure \ref{fig:wdtemp}). Note that in making this estimate we have not applied an extinction correction to these fluxes. As we noted in Section \ref{sec:xmmxray}, the galactic $E(B-V)$ is negligible, but we measure some extra absorption in the X-ray data. This is probably related to the high inclination of this system and due to our looking through a significant amount of gas above the accretion disc. Similar effects are seen in a number of CVs, such as V893~Sco \citep{Mukai09}. This material will not be in the form of dust and so will not cause `extinction' in the classical sense, but it is possible that it may cause an `accretion curtain' in this system, such as in OY~Car \citep{Horne94}. If this was the case then our temperature determination was a lower limit. However, it is impossible to detect such an effect without UV spectroscopic data. Our temperature determination for the white dwarf is consistent
(within the uncertainties) with the theoretical models of \citet{Bildsten06}, which predict a temperature of $18,000$K for this system.

\section{LIGHT CURVE ANALYSIS}
\label{sec:analysis}

In this section we describe the model we fitted to the phase-folded and binned WHT/ULTRACAM $2006$ and $2009$ data in order to make parameter determinations for this system. For the $2006$ data it was first necessary to subtract the night-to-night variations caused by the superhump; this is detailed in Section \ref{sec:superhump}. Secondly, we fitted the data on an eclipse-by-eclipse basis in order to refine the ephemeris, which we discuss in Section \ref{sec:ephemeris}. We then phase-folded the data using this ephemeris and fitted it with a Markov Chain Monte Carlo (MCMC) light-curve model in order to obtain final parameter determinations. We fit the three bands separately. This fitting process is described in Section \ref{sec:phasefold}.

\subsection{Determination of the ephemeris}
\label{sec:ephemeris}

\begin{figure*}
\centering
\includegraphics[angle=270,width=1.0\textwidth]{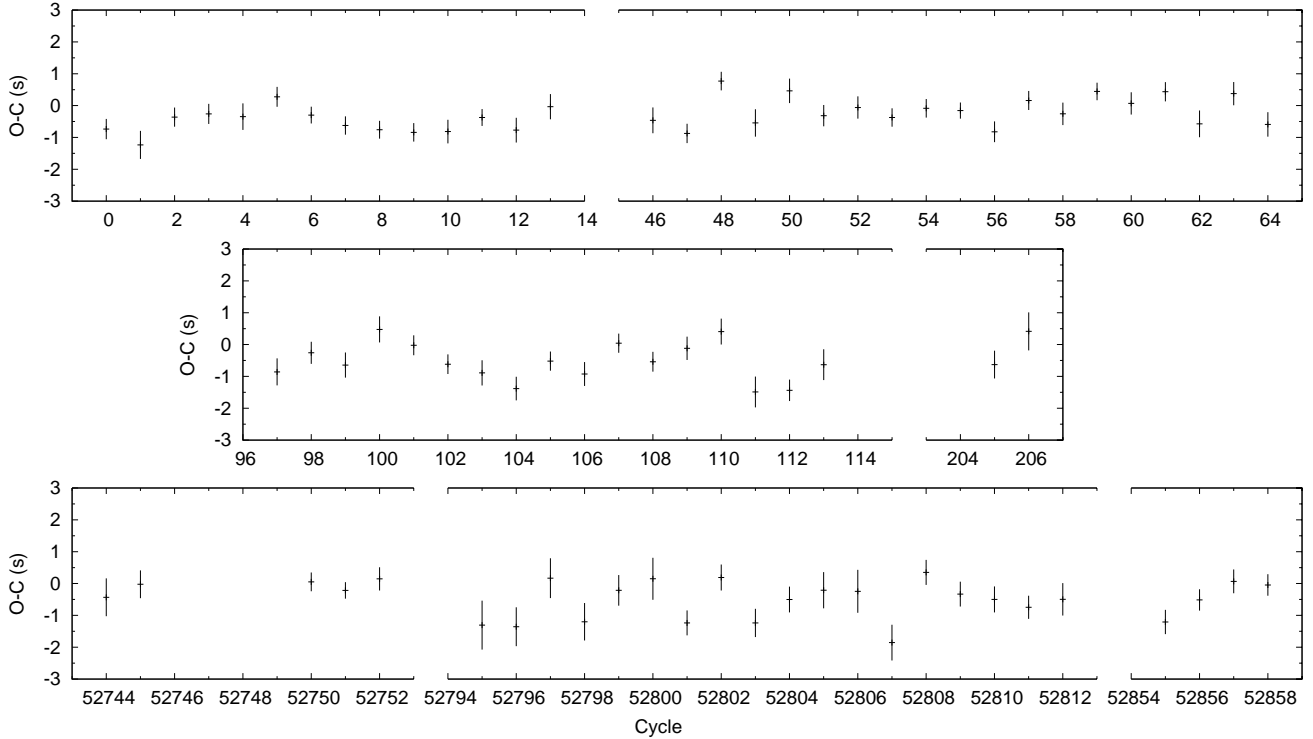}
\hfill
\caption{$(O-C)$ values plotted against cycle number, using the eclipse timings determined from the model fits to our WHT/ULTRACAM data and the linear ephemeris given in Section \ref{sec:ephemeris}. Cycle $0$ corresponds to the first eclipse in the $1$ March 2006 dataset. Note the break in the x-axis of this plot between the various nights of data.} \label{fig:ominusc} \end{figure*}

We divided each light curve into separate orbital cycles, and then fit the model determined in Section \ref{sec:phasefold} to each cycle using the Levendburg-Marquardt method \citep{Press02} in order to generate a series of eclipse timings. These timings are on the barycentric dynamical timescale, corrected for light travel to the solar system barycentre. A least-squares fit to all of these data yields the ephemeris \\

$MJD(TDB) = 53795.9455191(5) +  0.01966127289(2)E$ \\ 

\noindent for the mid-point of the white dwarf eclipse.  We plot the residuals of the linear ephemeris in Figure \ref{fig:ominusc}. There is some systematic variation, perhaps due to the residuals of the superhump or flickering. With only two epochs of observation any long-term departure from a linear ephemeris cannot be determined: a third epoch of high-speed observation will be necessary to identify any period changes in this system.

\subsection{Fitting the phase-folded light curves}
\label{sec:phasefold}

In order to make precise parameter determinations we chose to combine our data into phase-folded and binned light curves. We began by preparing the $2006$ data for fitting by subtracting the superhump modulation from the data, using the parameters determined in Section \ref{sec:superhump}. Since the superhump is not seen in $2009$, this step is not necessary for this second epoch of data. Once the superhump was subtracted it became clear that other binary parameters varied over the course of our observations, in particular the disc radius which changes significantly from night to night. These variations prevented us from combining our entire WHT/ULTRACAM dataset. We therefore chose to create separate phase-folded light curves for each individual night of observation. Even this could potentially introduce some systematic effect on our results due to the change in disc radius between the beginning and end of each night's observation. We examine this in detail in Section \ref{sec:discradius} and find that such effects are small, so we do not believe this influences our results to a significant degree. Since we cannot combine nights, we chose to omit the nights of $5$ March $2006$, $1$ January $2009$ and $3$ January $2009$, in which we only have a few cycles of data. We therefore fitted a total of twelve light curves: the $u'$-, $g'$- and $r'$-band phase-folded data for the nights of $1$, $2$ and $3$ of March $2006$; and $2$ January $2009$. 

We modelled the light curve with {\sevensize LCURVE}, a code developed to fit light-curves characteristic of eclipsing dwarf novae and detached white dwarf / M dwarf binary stars. A complete description of this code is given in the appendix of \citet{Copperwheat09}. We implemented this code in this work with two modifications, both to the bright spot component. In \citet{Copperwheat09} the bright spot is modelled as a line of elements which lie upon a straight line in the orbital plane. The surface brightness of the elements is parameterised with two power-law exponents. Since the bright spot in SDSS~0926 is a relatively weak component of the emission we do not require this degree of complexity. We therefore use a simpler bright spot model, setting the exponent $\gamma$ to $1$ (our bright spot model is therefore identical to the earlier prescription of \citealt{Horne94}). Secondly, in \citet{Copperwheat09} the angle $\phi$ was defined as the angle the line of elements of the bright spot makes with the line of centres between the two stars. We have changed the definition of this angle in our code: $\phi$ is now the angle the line of elements of the bright spot makes with the tangent to the outer edge of the accretion disc. This modification makes this angle easier to interpret in a physical context, since $\phi = 0$ implies a bright spot which runs along the outer edge of the disc, and so we would expect $\phi$ to be close to this. We noted in \citet{Copperwheat09} that the two angles $\phi$ and $\psi$ (the angle away from the perpendicular at which the light from the bright spot is beamed) are highly correlated and tend to be poorly constrained, and so here we fix $\phi=0$.

In the {\sevensize LCURVE} code the binary is defined by four components: a white dwarf primary, a Roche-lobe filling secondary star, accretion disc and bright-spot. We first obtained an initial fit to each light curve using the simplex and Levenberg-Marquardt methods \citep{Press02}. We then used a Markov Chain Monte Carlo (MCMC) algorithm for minimisation and determination of uncertainties (details of our MCMC method are also given in \citealt{Copperwheat09}). We fitted all of the parameters used in \citet{Copperwheat09}, with the exception of $\gamma$ and $\phi$, as described above. We additionally set the accretion disc radius $R_{disc}$ equal to $R_{spot}$: the distance between the bright spot and the primary white dwarf. We therefore assume that the `head' of the bright spot is on the outer rim of the disc. $R_{disc}$ is rather poorly constrained by our data, and as we showed in \citet{Copperwheat09} $R_{spot}$ is highly correlated with $R_{disc}$, so this is a reasonable approximation. One additional parameter which our results are potentially sensitive to is the limb darkening of the primary white dwarf. We made an initial fit to the data in order to estimate the effective temperature and surface gravity (we discuss these quantities in Section \ref{sec:components}), and then used a model atmosphere code (described in \citealt{Koester10}) to calculate the specific intensity at different points across the stellar disc. We then fitted these values to determine the limb darkening coefficients. We tried various limb darkening laws, and found the best fit was for the four-parameter law of \citet{Claret00}, although the choice between this and a fourth order polynomial is unlikely to influence our results. We list our determinations of the coefficients in Table \ref{tab:limbdark}. These values were used in all our MCMC fits.

\begin{table} 
\caption{Limb darkening coefficients for the primary white dwarf, using the four-coefficient law of \citet{Claret00}.}
\label{tab:limbdark} 
\begin{tabular}{lrrr} 
Coefficient  &$u'$ &$g'$ &$r'$\\
\hline
$1$         &$1.35$    &$1.18$   &$1.37$\\    
$2$         &$-1.30$  &$-1.13$  &$-2.07$\\
$3$         &$0.90$   &$0.75$   &$1.96$\\
$4$         &$-0.28$  &$-0.23$  &$-0.72$\\
\hline		
\end{tabular}
\end{table}

\begin{figure*}
\begin{center}
	\includegraphics[angle=270,width=1.0\textwidth]{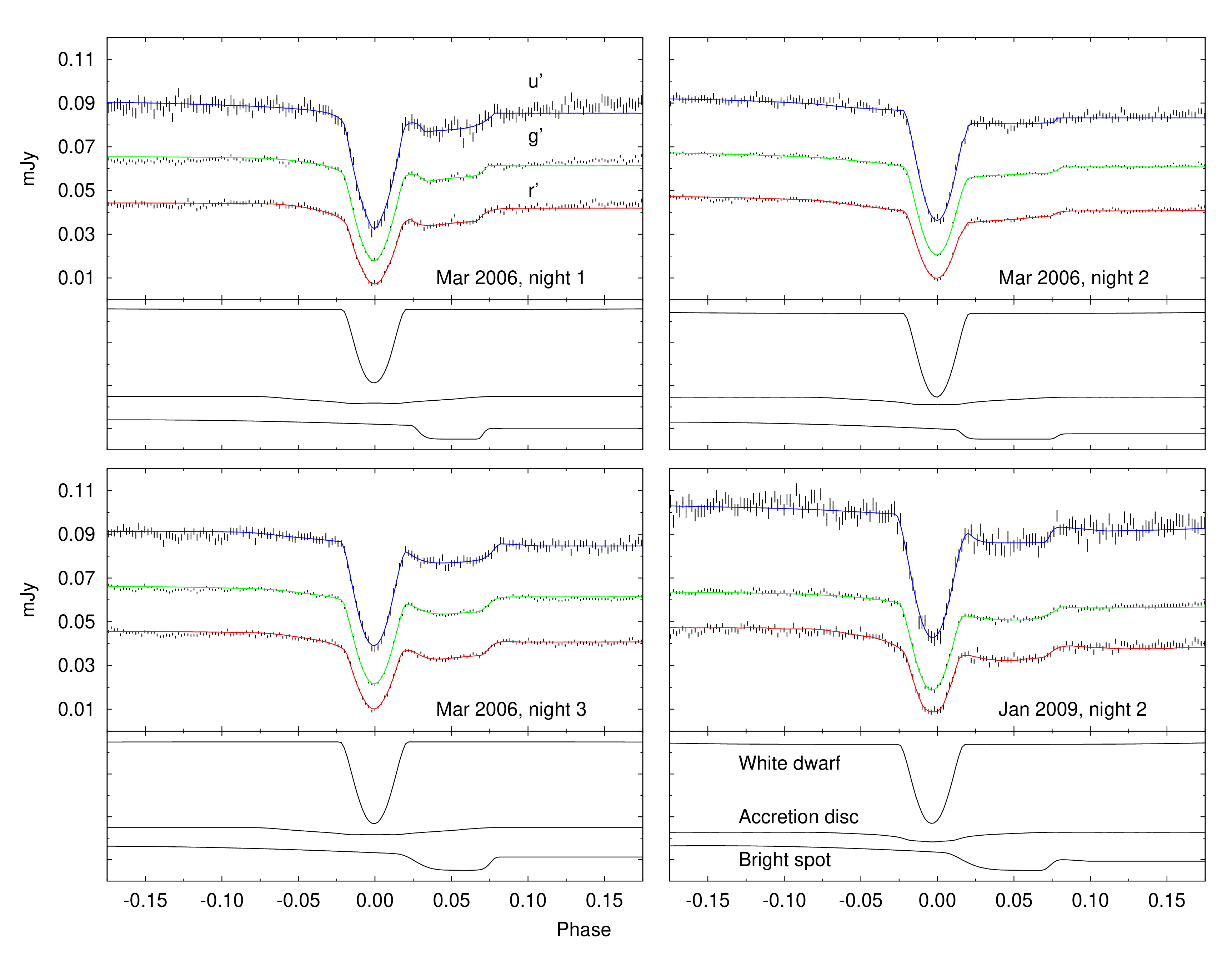}
\end{center}
\caption{Phase-folded and binned light curves for the first, second and third nights of the $2006$ observations, and the second night of the $2009$ observations. For each we plot the three bands separately (top, $u'$; middle, $g'$; bottom, $r'$). We plot the average flux in mJy against the binary phase, where a phase of $0$ corresponds to the mid-eclipse of the white dwarf. We plot the datapoints with uncertainties in black, and the best model fits to these data in blue, green and red (for $u'$, $g'$ and $r'$, respectively). For clarity we apply offsets of $0.005$mJy to the $g$-band data and $0.01$mJy to the $u$-band. For each night we also show underneath the lightcurve the three components of the $g'$-band model plotted separately, showing the relative strengths of the bright spot, accretion disc and white dwarf. These three lines have also been offset for clarity.} \label{fig:phasefolded_all}
\end{figure*}

The parameter determinations from these fits are listed in Table \ref{tab:mcmc}. Note that the uncertainties on these MCMC results are non-Gaussian, and so the values we quote in this table only provide an approximate description of the uncertainties. We plot the distribution of the mass ratio and inclination values from our fits in Figure \ref{fig:qvsi}. Additionally, in Figure \ref{fig:phasefolded_all} the phase-folded light curves for the four nights are plotted, along with the best model fits.  If we examine Figure \ref{fig:phasefolded_all} first, we see we obtain a good fit to the data in all three bands for each of the four nights. The dominant component in the light curve is the primary white dwarf, and the main eclipse in the light curve is the eclipse of this feature. Note that the eclipse is round-bottomed, meaning it is a partial eclipse of the white dwarf. The second, smaller eclipse in these light curves is of the bright spot, which is a much weaker component. The phase of this eclipse clearly varies with respect to the phase of the white dwarf eclipse: this can be understood in terms of a change in the relative position of the bright spot due to variations in the disc radius. Compare for example the first night of 2006 with the second. On the first night, the two eclipses are clearly distinct, with the egress of the white dwarf followed by the ingress of the bright spot. On the second night, the egress and the ingress overlap. This implies a larger apparent accretion disc radius on the second night, due to the precession of the elliptical disc. The two other nights of data lie somewhere between these two extremes. The fact that the two eclipses are sometimes separate implies a very low mass ratio for this system: in higher mass ratio systems we would expect to see the ingress of both the white dwarf and the bright spot before the white dwarf egress. 

We now turn to the parameter determinations listed in Table \ref{tab:mcmc}. The first four of these are the parameters that are important in characterising the system: these are the mass ratio $q$, the binary inclination $i$, the primary white dwarf radius scaled by the binary separation $R_1/a$, and the accretion disc radius scaled by the binary separation $R_{disc}/a$. The mass ratio and the inclination are highly correlated, and we discuss these parameters in Section \ref{sec:qandi}. The white dwarf temperature is used with $q$, $i$ and $R_1/a$ to derive the remaining binary parameters in Section \ref{sec:components}. We discuss the disc radius in more detail in Section \ref{sec:discradius}. The remaining six parameters in Table \ref{tab:mcmc} pertain to the accretion disc and the bright spot. We find these parameters to be rather poorly constrained in general, due to the fact that the disc and the bright spot make a relatively weak contribution to the flux in this system. However, we find no correlation between these parameters and the `important' parameters listed above, and so the uncertainty in these values does not imply any further systematic uncertainty in the determinations of Sections \ref{sec:qandi} and \ref{sec:discradius}.

\begin{table}
\caption{Results from our MCMC fits to the phase-folded light curves, as detailed in Section \ref{sec:phasefold}. $q$: the mass ratio. $i$: the binary inclination. 
$R_1$: the white dwarf radius. $R_{disc}$: the accretion disc radius. $\beta$: the power-law exponent for the bright spot. $\delta$: the exponent of surface brightness over the accretion disc. $l$: the bright spot scale-length. $f_c$: the fraction of the bright spot taken to be equally visible at all phases. $\phi$: the angle made by the line of elements of the bright spot, measured in the direction of binary motion from the line tangental to the accretion disc. $\psi$: the angle away from the perpendicular at which the light from the bright spot is beamed, measured in the same way as $\phi$. A complete description of all these parameters is given in the appendix of \citealt{Copperwheat09}}
\label{tab:mcmc} 
\begin{tabular}{lr@{\,$\pm$\,}lr@{\,$\pm$\,}lr@{\,$\pm$\,}l} 
\hline
        &\multicolumn{2}{c}{$u'$} &\multicolumn{2}{c}{$g'$} &\multicolumn{2}{c}{$r'$}\\
\hline       
&\multicolumn{6}{c}{2006, night 1}\\   \\  
$q$             &$0.037$ &$0.003$   &$0.051$ &$0.004$   &$0.043$ &$0.003$   \\
$i$             &$82.94$ &$0.29$    &$81.89$ &$0.25$    &$82.53$ &$0.23$    \\
$R_1/a$         &$0.037$ &$0.004$   &$0.036$ &$0.001$   &$0.027$ &$0.002$   \\
$R_{disc}/a$    &$0.341$ &$0.013$   &$0.336$ &$0.049$   &$0.339$ &$0.012$   \\
$\delta$        &$-1.86$ &$0.55$    &$-2.23$ &$0.39$    &$-1.99$ &$0.22$    \\
$l$             &$0.040$ &$0.006$   &$0.026$ &$0.002$   &$0.018$ &$0.004$     \\
$f_c$         	&$0.914$ &$0.018$   &$0.846$ &$0.024$   &$0.921$ &$0.012$     \\
$\beta$         &$2.04$ &$1.00$     &$2.51$ &$0.76$     &$2.25$  &$1.01$     \\
$\psi$          &$22.5$ &$15.2$     &$172.1$ &$12.7$    &$21.4$  &$7.7$            \\
\hline       
&\multicolumn{6}{c}{2006, night 2}\\\\   
$q$             &$0.038$ &$0.002$   &$0.046$ &$0.003$   &$0.042$ &$0.002$       \\ 
$i$             &$82.58$ &$0.22$    &$82.13$ &$0.28$    &$82.52$ &$0.19$        \\
$R_1/a$         &$0.040$ &$0.001$   &$0.034$ &$0.001$   &$0.035$ &$0.002$       \\
$R_{disc}/a$    &$0.455$ &$0.012$   &$0.415$ &$0.013$   &$0.435$ &$0.008$       \\
$\delta$        &$-0.71$ &$0.32$    &$-1.90$ &$0.21$    &$-0.80$ &$0.22$         \\
$l$             &$0.002$ &$0.003$   &$0.011$ &$0.003$   &$0.004$ &$0.002$         \\
$f_c$         	&$0.636$ &$0.014$   &$0.79$ &$0.02$     &$0.67$ &$0.02$         \\
$\beta$         &$1.54$	&$1.05$	    &$2.02$ &$0.97$     &$1.39$ &$1.11$         \\
$\psi$          &$46.8$ &$7.1$   	&$37.0$ &$6.3$      &$39.6$ &$9.9$      \\
\hline       
&\multicolumn{6}{c}{2006, night 3}\\ \\ 
$q$             &$0.039$ &$0.001$   &$0.039$ &$0.002$   &$0.037$ &$0.002$\\
$i$             &$82.77$ &$0.16$    &$82.80$ &$0.21$    &$82.89$ &$0.18$\\
$R_1/a$         &$0.036$ &$0.001$   &$0.037$ &$0.002$   &$0.032$ &$0.002$\\
$R_{disc}/a$    &$0.394$ &$0.009$   &$0.369$ &$0.004$   &$0.370$ &$0.004$\\
$\delta$        &$-0.68$ &$0.19$    &$-0.88$ &$0.28$    &$-1.40$ &$0.28$\\
$l$             &$0.069$ &$0.001$   &$0.026$ &$0.004$   &$0.024$ &$0.002$\\
$f_c$         	&$0.916$ &$0.005$   &$0.91$ &$0.01$     &$0.91$ &$0.01$\\
$\beta$         &$0.92$ &$0.29$     &$3.00$ &$0.62$     &$3.10$ &$0.55$\\
$\psi$          &$29.0$ &$13.6$     &$24.6$ &$8.2$      &$23.9$ &$7.3$\\
\hline
&\multicolumn{6}{c}{2009, night 2}\\ \\ 
$q$             &$0.028$ &$0.004$   &$0.040$ &$0.001$   &$0.042$ &$0.006$\\
$i$             &$84.04$ &$0.45$    &$82.71$ &$0.08$    &$82.80$ &$0.45$\\
$R_1/a$         &$0.043$ &$0.003$   &$0.032$ &$0.001$   &$0.015$ &$0.005$\\
$R_{disc}/a$    &$0.289$ &$0.012$   &$0.393$ &$0.002$   &$0.379$ &$0.019$\\
$\delta$        &$1.10$ &$1.27$     &$-1.89$ &$0.12$    &$-1.92$ &$0.17$\\
$l$             &$0.174$ &$0.006$   &$0.029$ &$0.002$   &$0.055$ &$0.015$\\   
$f_c$         	&$0.95$ &$0.01$     &$0.82$ &$0.01$     &$0.69$ &$0.12$\\
$\beta$         &$2.62$ &$0.74$     &$2.34$ &$0.51$     &$1.12$ &$2.76$\\
$\psi$          &$30.43$ &$12.1$    &$22.9$ &$3.4$      &$20.4$ &$27.0$\\
\hline
\end{tabular}
\end{table}

\section{DISCUSSION}
\label{sec:discussion}

\begin{table*} 
\caption{Binary parameters for SDSS~0926. We list the mass ratio $q$, inclination $i$, and primary radius $R_1$ as determined from the model fits. We derive the remaining binary parameters (the binary separation $a$, the component masses $M_1$ and $M_2$ and the radial velocity semi-amplitudes $K_1$ and $K_2$) from these values and our calculated mass/radius relation for the white dwarf (Section \ref{sec:components}). The donor radius $R_2$ is calculated using the approximation of \citet{Eggleton83}.}
\label{tab:params} 
\begin{tabular}{llr@{\,$\pm$\,}lr@{\,$\pm$\,}lr@{\,$\pm$\,}l} 
\hline
&&\multicolumn{2}{c}{$u'$-band}  &\multicolumn{2}{c}{$g'$-band}  &\multicolumn{2}{c}{$r'$-band}\\
\hline
$q$ &		                        &$0.038$ &$0.003$           &$0.041$ &$0.002$       &$0.040$ &$0.002$\\
$i$ &(deg)	                        &$82.8$ &$0.3$             	&$82.6$ &$0.3$          &$82.7$ &$0.2$\\	
$R_1 / a$ &(\Rsun)                  &$0.038$ &$0.003$         	&$0.033$ &$0.002$     	&$0.031$ &$0.005$\\
\hline
$a$ &(\Rsun)	                    &$0.281$ &$0.007$		    &$0.295$ &$0.005$	    &$0.299$ &$0.012$\\
$M_1$ &(\Msun)	                    &$0.74$ &$0.05$		        &$0.85$ &$0.04$		    &$0.90$ &$0.10$\\
$M_2$ &(\Msun)	                    &$0.028$ &$0.004$		    &$0.035$ &$0.003$	    &$0.036$ &$0.006$\\
$K_1$ &(km/s)	                    &$26$ &$3$			        &$30$ &$2$		        &$29$ &$3$\\
$K_2$ &(km/s)	                    &$692$ &$18$		        &$723$ &$13$	    	&$735$ &$29$\\
$R_2$ &(\Rsun)	                    &$0.044$ &$0.002$		    &$0.047$ &$0.001$	    &$0.047$ &$0.003$\\
\hline
\end{tabular}
\end{table*}

\subsection{Mass ratio and inclination}
\label{sec:qandi}

\begin{figure*}
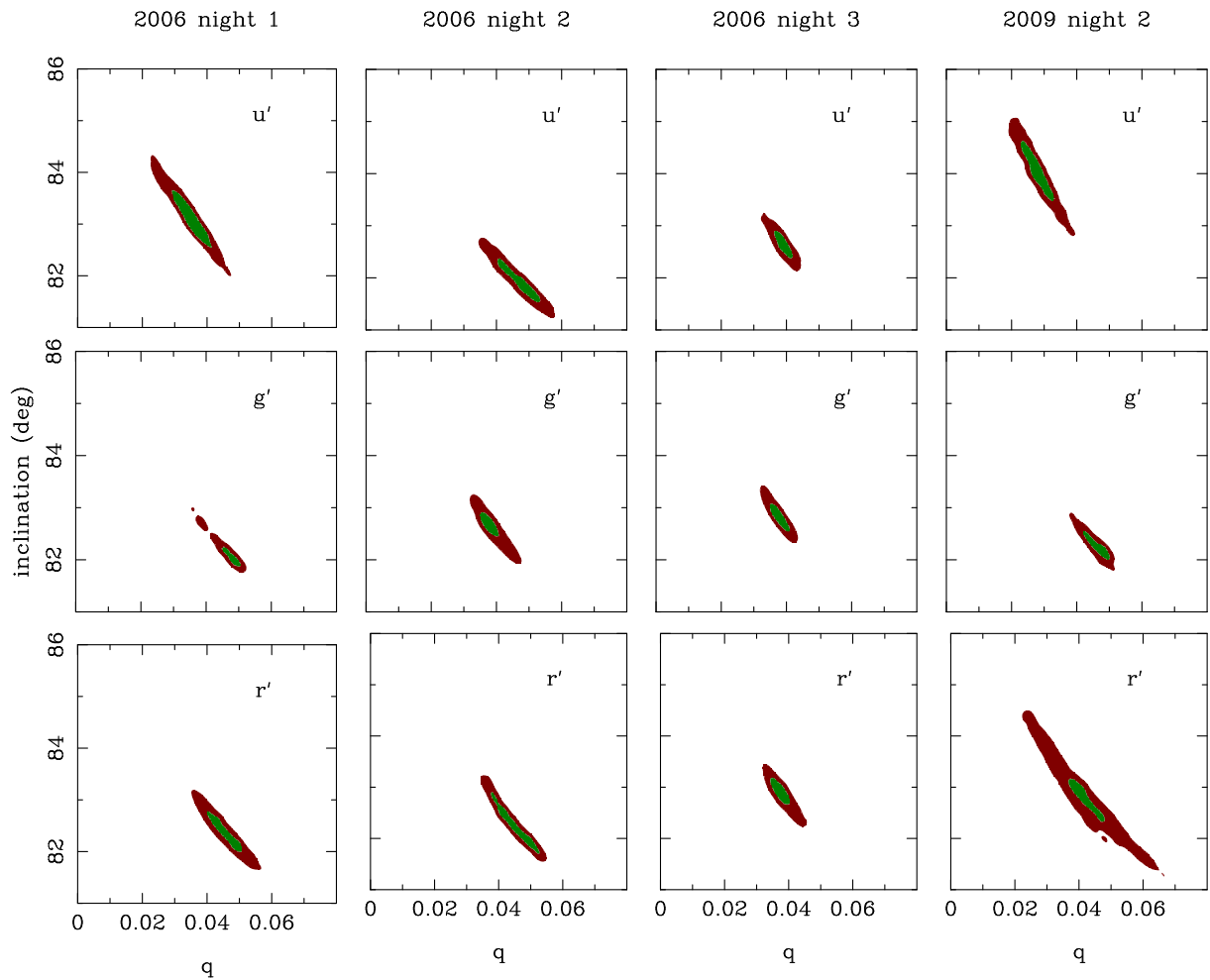

\begin{center}$
\begin{array}[t]{rccc}
    \includegraphics[angle=270,width=0.22\textwidth]{2006_u_1.ps} & \includegraphics[angle=270,width=0.2\textwidth]{2006_u_2.ps} & \includegraphics[angle=270,width=0.2\textwidth]{2006_u_3.ps} & \includegraphics[angle=270,width=0.2\textwidth]{2009_u_2.ps}\\
   \includegraphics[angle=270,width=0.2\textwidth]{2006_g_1.ps} & \includegraphics[angle=270,width=0.2\textwidth]{2006_g_2.ps} & \includegraphics[angle=270,width=0.2\textwidth]{2006_g_3.ps} & \includegraphics[angle=270,width=0.2\textwidth]{2009_g_2.ps}\\   
    \includegraphics[angle=270,width=0.22\textwidth]{2006_r_1.ps} & \includegraphics[angle=270,width=0.2\textwidth]{2006_r_2.ps} & \includegraphics[angle=270,width=0.2\textwidth]{2006_r_3.ps} & \includegraphics[angle=270,width=0.2\textwidth]{2009_r_2.ps}\\     
\end{array}$
\end{center}
\caption{The distribution of the mass ratio $q$ and inclination $i$ results from our MCMC runs, with the green and red regions indicating the $1\sigma$ and $3\sigma$ confidence intervals. We plot (from left to right) nights 1, 2 and 3 of the 2006 observations, and night 2 of the 2009 observations. For each night we plot (top to bottom) the distribution in the $u'$, $g'$ and $r'$ bands.} \label{fig:qvsi}
\end{figure*}

If we assume a Roche-lobe filling donor star, the phase width of the white dwarf eclipse is then an observable quantity that is intrinsically linked to two physical properties: the mass ratio and the binary inclination. For a higher binary inclination the duration of the eclipse will be greater, thus to maintain the same phase width, as the inclination is increased, the size of the donor, and hence the mass ratio, must be decreased. There is therefore a unique relationship between these two properties \citep{Bailey79}. This degeneracy can be broken since we have an additional geometric constraint due to the ingress and egress of the bright spot. The path of the accretion stream and hence the position of the bright spot is modified by the mass ratio. With this additional information we can determine both the mass ratio and inclination in this system. 

Our MCMC results for the mass ratio versus inclination are plotted in Figure \ref{fig:qvsi}. We see that these points are distributed along a curved path across this plot: this is the line of constant phase width for the white dwarf, which we measure to be $\sim$$0.0220$. The points are constrained to this line, with the scatter due to the uncertainty in the phase width. We calculate the weighted means of the $q$, $i$ and $R_1 / a$ values in Table \ref{tab:mcmc}, and present the results in Table \ref{tab:params}. We emphasise at this point that the weighted means may underestimate the uncertainties in these parameters. The MCMC results in Table \ref{tab:mcmc} show variations which are formally significant, most likely due to effects such as flickering and residuals from the superhumps. In addition, the bright spot component is weak in this system, and so the best fit model for this component can be quite different in different bands, particularly in $u'$. It is important therefore to note that the results we discuss in this and the following sections, and in Table \ref{tab:params}, may not fully account for these systematic effects, and the magnitude of these effects are better understood by referring to the individual night's results in Table \ref{tab:mcmc}.

We find the mass ratio $M_2/M_1$ to be $q = 0.041 \pm 0.002$ in our $g'$-band measurements. The inclination is $i =82.6 \pm 0.3$ deg in $g'$. The values of $q$ and $i$ in the other two bands are consistent with these values. Our $2006$ data were originally published in \citet{Marsh07} and there we reported different values of $0.035 \pm 0.002$ and $83.1 \pm 0.1$ deg for $q$ and $i$. This discrepancy is due to a number of factors. Primarily, the MCMC method we use for minimisation is superior to the Levenberg-Marquardt minimisation employed in \citet{Marsh07}. We have found in situations of strong degeneracy, such as between $q$ and $i$ here, the Levenberg-Marquardt and simplex methods have a tendency to stop before finding the true minimum \citep{Copperwheat09}. Additionally, we have used more appropriate limb-darkening coefficients in this work, since we have been able to estimate the white dwarf temperature. \citet{Marsh07} also arrived at their parameter determinations by combining data from different nights, which as we have discussed may introduce some systematic uncertainty due to the changing disc radius.

\subsection{Other binary parameters}
\label{sec:components}

In Table \ref{tab:params} we list our determination of the primary white dwarf radius, scaled by the binary separation. We used this, combined with our determination of the mass ratio and the binary inclination, to calculate the remaining binary parameters. One additional piece of information that was needed for this is a mass/radius relation for the primary white dwarf. 

We determined this by using the white dwarf temperature, which we found in Section \ref{sec:xmmuv} to be $17,000$K. However this is an iterative process, since the temperature determination itself requires a mass/radius relation. The process was as follows. We began by assuming a mass/radius relation for the white dwarf, and for this we used the Eggleton zero-temperature mass/radius relation (quoted in \citealt{Verbunt88}). We subsequently determined the mass and radius of the white dwarf, and from our model calculated the white dwarf contribution in each band. It is these white dwarf fluxes which we list in Section \ref{sec:xmmuv}. We then used the white dwarf model atmospheres of \citet{Gaensicke95} to calculate the white dwarf temperature $T_{eff}$, fixing $\log g$ to the value implied by the zero-temperature mass and radius. We then compared these values of $T_{eff}$ and $M_1$ to the white dwarf cooling models of \citet{Bergeron95} in order to find the white dwarf radius which is consistent with these values. By comparing this to the Eggleton zero-temperature radius, we determined an `oversize factor' for the white dwarf, which we found to be $1.03$. Finally, we determine a final white dwarf mass and radius using a new mass/radius relation, which is the zero-temperature relation scaled by this oversize factor. In theory, at this point we should then re-iterate this process and use the new mass and radius to refine the temperature determination. However, since the first oversize factor is very close to $1$ the effect of further iterations is negligible: the white dwarf temperature fit is not affected by the very small change in $\log g$ which results from a $3 \%$ increase in the white dwarf radius. Note also that since the oversize factor is so close to $1$, while the uncertainty in our temperature measurement may be quite large (since it is based on a small number of flux measurements) it will not affect our parameter determinations to a significant degree. 

In the second half of Table \ref{tab:params}, we list the binary separation $a$, the masses of the two components, the radial velocity semi-amplitudes of the two components and the radius of the secondary. This radius is the volume radius of the Roche lobe filling donor, which we calculate using the approximation of \citet{Eggleton83}. In our $g'$-band fits, we find the mass of the primary white dwarf to be $0.85 \pm 0.04$\Msun. The mass of the donor is $0.035 \pm 0.003$\Msun \ and the donor radius is $0.047 \pm 0.001$\Rsun. In \citet{Marsh07}, we reported the primary mass to be $0.84 \pm 0.05$\Msun \ and the donor mass to be $0.029 \pm 0.002$\Msun. These values are close to our updated findings -- the differences are primarily due to the reasons discussed in Section \ref{sec:qandi}, as well as our accounting for the finite temperature of the primary white dwarf in this work.

Finally, by using our measurements of the white dwarf contribution, along with the theoretical absolute magnitudes from the \citet{Bergeron95} and \citet{Holberg06} cooling models, we can estimate the distance modulus for this system. Using the absolute magnitudes for a white dwarf temperature of $17,000$K and mass of $0.8$\Msun, our flux measurements imply a distance of $460$ -- $470$pc.

\subsection{AM CVn formation scenarios}
\label{sec:formation}

A donor mass of $0.035 \pm 0.003$\Msun \ implies the donor is only partially degenerate: a fully degenerate donor in a system with this period would have a mass of $\sim$$0.020$\Msun. \citet{Roelofs07b} measured the masses of five systems using parallax measurements obtained with \hst, and found four of the five to be consistent with a partially degenerate donor. The degree of degeneracy varied, but the donor was typically found to have a mass of between $2$ and $4$ times the mass of a fully degenerate donor. The lowest estimate was for HP~Lib, which was found to have a donor mass of $1.6$ -- $2.9$ times the fully degenerate mass. At $1.75$ times the fully degenerate mass, SDSS~0926 is by comparison at the lower end of these estimates.

We now examine the finding of a donor mass of $0.035 \pm 0.003$\Msun \ in the context of the evolutionary history of this system. There are three proposed formation paths for AM CVn binaries and all three are consistent with a donor that is partially degenerate to some degree. The `white dwarf channel' \citep{Nelemans01} suggests detached close double white dwarfs which are brought into contact as a result of angular momentum loss due to gravitational wave radiation (GWR). \citet{Nelemans01} used a zero-temperature donor in their formulation, but \citet{Deloye05} argued that the donors could be semi-degenerate to some degree, depending on the contact time of the binary. The second formation path is the `helium star channel' \citep{Iben91}. In this scenario the donor is a low-mass, non-degenerate helium burning star. Following contact, material is accreted from the helium star onto the primary white dwarf until a donor mass of $\sim 0.2$\Msun \ is reached, at which point core helium burning ceases and the star becomes semi-degenerate. Further mass transfer driven by GWR sees the donor mass decrease and the orbital period increase to values consistent with the observed AM CVn population. The third scenario is the `evolved CV channel' \citep{Podsi03} which suggests the progenitors of AM CVns are CVs with evolved secondaries. The donor in this channel is initially non-degenerate and hydrogen-rich, but becomes degenerate and helium-rich (but still with a few per cent hydrogen) during its evolution before Roche lobe overflow.

\citet{Nelemans01} approximated the mass-radius relationship for a helium star donor and thus modelled the evolution of AM CVns formed via the helium star channel. Based on the examples provided, an AM CVn with a period equal to that of SDSS~0926 should have a donor mass of $\sim$$0.05$\Msun \ if it were formed by this channel. Table 1 of \citet{Podsi03} lists model parameter determinations for six AM CVns assuming they were formed via the evolved CV channel. For the two systems in this table closest in period to SDSS~0926 (V803 Cen and CP Eri), these results predict the donor mass to be $\sim$$0.04$\Msun \ at the point at which the orbital period of the system begins to increase. 

Since the donor mass in \citet{Marsh07} was found to be $0.029 \pm 0.002$\Msun, lower than the theoretical values proposed for the helium star and evolved CV channels, \citet{Deloye07} argued that this was evidence for formation via the white dwarf channel. The updated donor mass value we report here of $0.035 \pm 0.003$\Msun \ is consistent with formation via the evolved CV channel, and is within a few $\sigma$ of the \citet{Nelemans01} value for the helium star channel. We conclude therefore that our current findings do not strongly preclude any of the three formation channels. More precise determinations are likely to be possible if we are able to make more observations of SDSS~0926 in its non-superhumping state, since even after subtracting the superhump from our data there is likely to still be some residual systematic effect. The discovery of new eclipsing AM CVn systems is also key, particularly at the short end of the period distribution where there is the biggest discrepancy between the various donor mass predictions.

\subsection{The disc radius}
\label{sec:discradius}

\begin{figure}
\centering
\includegraphics[angle=270,width=1.0\columnwidth]{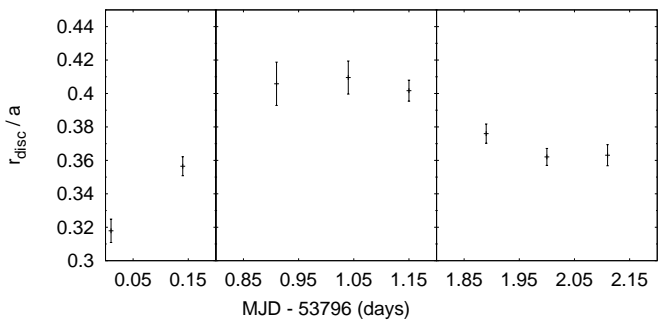}
\hfill
\caption{Variation in the accretion disc radius over the course of our 2006 observations. We divide the three nights of data into groups of five or six orbital cycles, which we phase-fold and fit. We plot here the accretion disc radius $R_{disc}$ scaled by the binary separation $a$, against the MJD of the mid-point of each block of data. The three panels of the plot show the results for the first, second and third night.} \label{fig:discradius} \end{figure}

In Table \ref{tab:mcmc} we list the apparent accretion disc radius for each night as determined from our MCMC fits. This is determined from the position of the bright spot, since we assume the bright spot to be at the outer edge of the accretion disc. We see in this table that the disc radius changes by between $10$ and $20$\% from night to night. As we discussed in Section \ref{sec:phasefold} this is also apparent in the lightcurves (Figure \ref{fig:phasefolded_all}) with the phase of the bright spot eclipse changing between nights. 

The changes we observe here are not due to radial variations in a circular disc, rather the disc is non-circular and the measured positions of the bright spot sample the possible range in disc radii. The superhumps we observe are generally taken to imply an elliptical and precessing disc, but the disc shape may be more complex than this, with detailed numerical simulations suggesting an irregular disc shape in superhumping AM CVn systems \citep{Simpson98}. The changing bright spot location we find in this system is in strong contrast with the study of AM CVn itself by \citet{Roelofs06}, who found very little variation in the bright spot position. This difference may be related to the very different \Mdot s which would be expected in these two systems. However, AM CVn was also found to be inconsistent with the Patterson $\epsilon$ -- $q$ relation (Section \ref{sec:superhump}), which is presumed to be independent of \Mdot.

We investigated our bright spot findings further by dividing our data into sections, in order to see if the disc radius variations can be observed over the course of a single night. We split each night of the $2006$ data into sections that are $5$/$6$ orbital cycles in length, and phase-folded and fitted each section individually (Figure \ref{fig:discradius}). We find there is a noticeable increase in radius when we compare the two halves of the first night, with $R_1 / a$ increasing from $0.318 \pm 0.007$ to $0.356 \pm 0.006$. During the second night the disc radius appears to be approximately constant over the course of our $8$h observation, and over the course of the third night a slight decrease in radius is observed when the two halves of the night are compared. These variations appear consistent with the precession period of the superhump.

\section{CONCLUSIONS}
\label{sec:conclusions}

In this paper we have presented high time resolution observations of the eclipsing $P = 28$ min AM CVn binary SDSS~0926 obtained with the fast CCD camera ULTRACAM mounted on the William Herschel Telescope. The primary aim of these observations was precise parameter determinations for the two binary components, using the photometric method. We determine the mass ratio to be $q = M_2 / M_1 = 0.041 \pm 0.002$, and the inclination to be $82.6 \pm 0.3$ deg. We calculate the mass of the primary white dwarf to be $0.85 \pm 0.04$\Msun, and find the donor to be partially degenerate with a mass of $0.035 \pm 0.003$\Msun. We also measure the eclipse timings with precision, and should be able to detect the period change due to gravitational wave losses with a third epoch of high time resolution observations.

We observed in our $2006$ WHT/ULTRACAM data the superhump that has previously been reported in this source. We determine the period of this variation and find it to be in agreement with the period excess / mass ratio relationship proposed by \citet{Patterson05}. This phenomenon is not present in the $2009$ WHT/ULTRACAM data, but we do see evidence for it in some LT/RISE data collected over the first half of $2009$. In addition to the superhump we observe a quasi-periodic oscillation in the 2006 data, with a period of around $50$s. Another interesting feature of our data is that we observe the accretion disc radius to be highly variable. We ascribe this to the tidal instability of the outer disc and observe changes of up to $10$\% in the radius over successive nights.

We obtained X-ray and ultraviolet observations with \xmm. We found no clear evidence for an eclipse in the X-ray light curve. The eclipse is detected in the ultraviolet with the \xmm \ Optical Monitor, and by fitting our model to these data we inferred the ultraviolet fluxes of the primary white dwarf. We used these along with our ground-based determinations of the white dwarf optical colours to determine the temperature of the white dwarf, which we found to be $17,000$K.

Using data collected with the Catalina Real-Time Transient Survey, we examined the outbursting behaviour of SDSS~0926 over a four and a half year period. We observed six outbursts over this period, in which the source flux increases by $\sim$$2$ -- $2.5$ magnitudes. The average time between outbursts is $\sim$$100$ - $200$ days. We observed in detail one additional outburst with LT/RISE. These data show an increase in source flux of greater than a factor of $3$, and a rapid decay. A lightcurve obtained one day after the detection of the outburst shows a complex structure, with the primary eclipse suggesting an asymmetrical distribution of disc flux or possibly an enhanced bright spot emission. There is additionally a dimming prior to the main eclipse which we suggest may be due to a warp in the disc itself.

\section*{ACKNOWLEDGEMENTS}
CMC and TRM are supported under grant ST/F002599/1 from the Science and Technology Facilities Council (STFC). ULTRACAM, VSD and SPL are supported by STFC grants PP/D002370/1 and PP/E001777/1. DS acknowledges the support of an STFC Advanced Fellowhip. The results presented in this paper are based on observations made with the William Herschel Telescope operated on the island of La Palma by the Isaac Newton Group in the Spanish Observatorio del Roque de los Muchachos of the Institutio de Astrofisica de Canarias; observations made with the Liverpool Telescope operated on the island of La Palma by Liverpool John Moores University in the Spanish Observatorio del Roque de los Muchachos of the Instituto de Astrofisica de Canarias with financial support from the UK Science and Technology Facilities Council; and on observations obtained with XMM-Newton, an ESA science mission with instruments and contributions directly funded by ESA Member States and NASA. The CSS survey is funded by NASA under grant NNG05GF22G issued through the Science Mission Directorate Near-Earth Objects Observations Program. This research has made use of NASA's Astrophysics Data System Bibliographic Services and the SIMBAD data base, operated at CDS, Strasbourg, France.

\bibliography{0926}

\end{document}